\documentclass[sigconf,nonacm]{acmart}
\usepackage[linesnumbered,ruled,vlined]{algorithm2e}
\usepackage{amsmath} % For \textnormal
\usepackage{multirow}
\usepackage{makecell}

\usepackage{amssymb}
\usepackage[dvipsnames]{xcolor}

\definecolor{MySemanticBlue}{rgb}{0.8392, 0.8823, 0.9372}
\definecolor{MyGeoGreen}{rgb}{0.7568, 0.8353, 0.6902}
\definecolor{MyRRFYellow}{rgb}{0.9568, 0.8588, 0.5255}
\AtBeginDocument{%
  }

\setcopyright{acmlicensed}
\copyrightyear{2026}
\acmYear{2026}
\acmDOI{XXXXXXX.XXXXXXX}
\begin{document}

%%
%% The "title" command has an optional parameter,
%% allowing the author to define a "short title" to be used in page headers.
% \title{Multimodal and Multiscale Spatial-Temporal Semantic Search and Recommendation with Large Vision-Language Models}
\title{Multimodal and Multiscale Spatial-Temporal Semantic Search and Recommendation with AI Foundation Models}

\author{Yuanyuan Tian}
\affiliation{%
  \institution{School of Geographical Sciences and Urban Planning, Arizona State University}
  \city{Tempe}
  \state{Arizona}
  \country{USA}}

\author{Wenwen Li} % Add a symbol next to the author's name
\affiliation{%
  \institution{School of Geographical Sciences and Urban Planning, Arizona State University}
  \city{Tempe}
  \state{Arizona}
  \country{USA}}

\authornote{For correspondence, please contact Dr. Wenwen Li (wenwen@asu.edu).}

\author{Xiao Chen}
\affiliation{%
  \institution{School of Geographical Sciences and Urban Planning, Arizona State University}
  \city{Tempe}
  \state{Arizona}
  \country{USA}}

\author{Michael Brook}
\affiliation{%
  \institution{Alaska Native Tribal Health Consortium}
  \city{Anchorage}
  \state{Alaska}
  \country{USA}}

\author{Michael Brubaker}
\affiliation{%
  \institution{Alaska Native Tribal Health Consortium}
  \city{Anchorage}
  \state{Alaska}
  \country{USA}}

\author{Anna Liljedahl}
\affiliation{%
  \institution{Woodwell Climate Research Center}
  \city{Falmouth}
  \state{Massachusetts}
  \country{USA}}

\author{Chitta Baral}
\affiliation{%
  \institution{School of Computing and Augmented Intelligence, Arizona State University}
  \city{Tempe}
  \state{Arizona}
  \country{USA}}

%%
%% By default, the full list of authors will be used in the page
%% headers. Often, this list is too long, and will overlap
%% other information printed in the page headers. This command allows
%% the author to define a more concise list
%% of authors' names for this purpose.
\renewcommand{\shortauthors}{Tian et al.}

%%
%% The abstract is a short summary of the work to be presented in the
%% article.
\begin{abstract}
Semantic search and recommendation of similar documents, such as news and reports about unusual environmental events (e.g., a dead whale washed ashore in Alaska) that contain spatial and temporal information, is a critical task in Geographic Information Retrieval (GIR). This work presents a novel framework that leverages AI foundation models, including Large Language Models (LLMs) and Vision-Language Models (VLMs), to enable effective similarity search and ranking for such event documents. To support this goal, we introduce two new strategies: (1) CAMERA (Context-Aware Multimodal Event Retrieval Algorithm), which fuses textual and visual information to generate richer embeddings than those derived from text alone; and (2) ASTRA (Adaptive Spatial and Temporal Re-ranking Algorithm), which improves similarity ranking by incorporating scale-dependent spatiotemporal relevance alongside semantic similarity. Experimental results, using a dataset from the Local Environmental Observer Network, demonstrate that our VLM-enhanced methods outperform unimodal, LLM-based approaches in similarity ranking effectiveness. By automatically linking relevant event reports, the proposed framework helps both data curators and the general public gain deeper insights into environmental change and its localized impacts. These findings highlight the potential of AI foundation models to advance GIR through multifaceted, intelligent analysis that integrates key geographic concepts: space, time, scale, and semantics.
\end{abstract}


%%
%% Keywords. The author(s) should pick words that accurately describe
%% the work being presented. Separate the keywords with commas.

% \received{20 February 2025}
% \received[revised]{12 March 2009}
% \received[accepted]{5 June 2009}

%%
%% This command processes the author and affiliation and title
%% information and builds the first part of the formatted document.
\maketitle

\section{INTRODUCTION}

Environmental monitoring increasingly relies on large volumes of heterogeneous reports describing unusual environmental events, such as wildlife mortality, extreme weather anomalies, ecosystem disruptions, and other indicators of environmental change \cite{baker2012local,ensor2015social}. These reports are often documented in community-based observation systems, scientific platforms, and news media, providing valuable situational awareness for environmental monitoring, disaster response, and environmental research. An important capability for such systems is the ability to discover and recommend events that are similar to a given query event across space and time \cite{seneviratne2021weather}. Identifying related events enables analysts to uncover emerging environmental patterns, track the spread of ecological phenomena, and better understand localized impacts of global environmental change \cite{ganguly2008data}.

In Geographic Information Retrieval (GIR), event similarity is inherently multidimensional. Environmental events are characterized by textual descriptions and their spatial location, temporal occurrence, and often visual evidence such as photographs or maps. Consequently, effective retrieval requires jointly reasoning over semantic content, geographic proximity, and temporal dynamics. Traditional retrieval methods, such as keyword-based approaches (e.g., BM25, TF-IDF), rely primarily on lexical overlap between documents. Although these techniques are efficient and interpretable, they struggle to capture deeper semantic relationships across heterogeneous reports. Recent neural retrieval approaches improve semantic matching by embedding queries and documents into a shared dense vector space using pretrained language models. However, these approaches typically treat documents as purely textual objects and overlook the geographic characteristics that fundamentally shape environmental events \cite{de2025report}.

Beyond semantic understanding, environmental event retrieval presents a unique challenge related to \emph{scale}. The spatial and temporal extent of environmental phenomena can vary dramatically depending on the event type and context. For example, a localized landslide may affect only a small geographic area over a short time window, whereas a mega wildfire or tsunami may span hundreds of kilometers and persist for months. Consequently, determining whether two events are related requires reasoning about the appropriate spatial and temporal scales at which similarity should be evaluated. Existing retrieval and ranking methods typically rely on fixed spatial or temporal thresholds, which may either exclude relevant events or include irrelevant ones when applied across heterogeneous environmental contexts. Addressing this scale dependency remains a longstanding challenge in GIR and GeoAI \cite{li2024geoai, janowicz2020geoai, li2020geoai}.

Recent advances in foundation models offer new opportunities for addressing these challenges. Large Language Models (LLMs) demonstrate strong capabilities in semantic reasoning, and Vision-Language Models (VLMs) enable multimodal understanding. 
Recent work has assessed the capabilities of LLMs in understanding complex geometries and topological spatial relations, highlighting both their potential and current limitations in geospatial reasoning \cite{ji2025foundation}. VLMs have been applied to interpret geospatial imagery through language-guided prompts, supporting environmental monitoring and analysis \cite{kadiyala2024implementation}.
These models have shown promise in various retrieval and recommendation tasks.
However, most existing applications of foundation models focus on representation learning or text generation, without explicitly incorporating geographic reasoning. In particular, two key challenges remain underexplored for environmental event retrieval. First, effectively integrating visual information with textual event descriptions remains difficult, especially when visual data needs to be interpreted in terms of geographical or environmental context. Second, existing retrieval systems rarely exploit the reasoning capabilities of LLMs to model the scale-dependent nature of spatiotemporal relevance.

To address these challenges, we propose a novel framework for multimodal and multiscale spatial-temporal event retrieval and recommendation. Our framework integrates foundation models into a two-stage architecture designed to bridge semantic understanding and geographic reasoning. In the first stage, we introduce the \textbf{Context-Aware Multimodal Event Retrieval Algorithm (CAMERA)}, which constructs enriched event representations by extracting structured visual cues from event images using a VLM. Rather than relying on black-box multimodal embeddings, CAMERA converts visual observations into concise textual descriptors capturing key aspects such as event category, geographic context, and seasonal characteristics. These cues are then integrated with textual metadata to produce multimodal-informed embeddings that remain compatible with standard dense and sparse retrieval pipelines. By combining dense semantic retrieval with lexical matching through rank fusion, CAMERA generates a high-recall candidate set of potentially related events.

In the second stage, we introduce the \textbf{Adaptive Spatio-Temporal Re-ranking Algorithm (ASTRA)}, which refines the candidate list by explicitly modeling the scale-dependent nature of environmental events. ASTRA leverages the reasoning capabilities of a LLM to infer query-specific spatial and temporal bandwidths that represent the effective impact scale of the query event. These inferred scales are used to transform raw geographic and temporal distances into adaptive similarity kernels, enabling the ranking model to adjust its sensitivity to spatial and temporal proximity dynamically. The resulting features are integrated within a supervised Learning-to-Rank framework that jointly considers semantic similarity, categorical alignment, and adaptive spatiotemporal relevance. By replacing fixed thresholds \cite{tian2025advancing} with query-dependent adaptive scales, ASTRA enables a more flexible and context-aware ranking strategy.

We evaluate the proposed framework using a large-scale dataset from the Local Environmental Observer (LEO) Network, which contains thousands of environmental event reports contributed by community observers and domain experts. Experimental results demonstrate that our approach improves retrieval and recommendation performance compared with both traditional information retrieval methods and recent neural re-ranking approaches. In particular, our results highlight the importance of integrating multimodal context and scale-aware geographic reasoning for environmental event discovery.

The contributions of this work are summarized as follows:

\begin{itemize}
\item We propose \textbf{CAMERA}, a multimodal retrieval framework that integrates structured visual cue extraction from VLM with hybrid dense and sparse retrieval for environmental event search.
\item We introduce \textbf{ASTRA}, a scale-aware re-ranking algorithm that leverages LLM reasoning to infer adaptive spatial and temporal scales for event similarity.
\item We demonstrate that combining multimodal representations with adaptive spatiotemporal reasoning significantly improves the effectiveness of spatial-temporal event retrieval.
\item We provide a comprehensive evaluation on a real-world environmental event dataset, highlighting the potential of foundation models to advance geographic information retrieval.
\end{itemize}

These contributions highlight a new direction for GeoAI systems in which foundation models are used for representation learning and for structured reasoning over geographic context and scale.

\section{RELATED WORK}

\paragraph{Geographic Information Retrieval.}
GIR extends traditional information retrieval by incorporating geographic context as an essential relevance factor \cite{purves2011geographic}. GIR systems often operate over heterogeneous data sources, including geographic databases, maps, scientific literature, and social media streams \cite{acheson2021extracting,middleton2018location,watanabe2018newsmap}. Geographic information can be explicitly stored in structured metadata, and also frequently appears implicitly in unstructured text through place names or descriptive spatial references. Therefore, extracting geographic meaning from text requires geoparsing and geocoding techniques that identify toponyms and link them to geographic coordinates using gazetteers or spatial knowledge bases \cite{purves2018geographic}. Natural language processing methods further support the extraction of spatial entities, place relations, and geographic context from text corpora \cite{janowicz2022know,shao2017extraction}. To incorporate geographic relevance into retrieval, many GIR systems combine textual similarity with spatial proximity measures, such as geographic distance or spatial overlap, within ranking functions \cite{martins2010learning}. Ontologies and gazetteers are also frequently used to support place disambiguation, query expansion, and thematic similarity computation \cite{hu2015metadata,janowicz2011semantics}. Although these approaches provide structured mechanisms for geographic reasoning, they often struggle with noisy or unstructured text and typically rely on fixed spatial or temporal buffers \cite{tian2025advancing}, limiting their ability to adapt to heterogeneous geographic phenomena.

\paragraph{Multimodal Retrieval.}
Multimodal retrieval improves document understanding by integrating multiple information modalities, most commonly text and images. Recent progress in VLMs has enabled large-scale multimodal representation learning through joint embedding spaces that align visual and textual content. A example model is CLIP \cite{radford2021learning}, which learns shared image-text representations via contrastive training and has become a widely adopted baseline for multimodal search tasks. Additional research has explored multimodal fusion strategies that combine visual and textual features through early fusion, feature concatenation, or late fusion mechanisms \cite{liang2024foundations}. Although joint embeddings capture cross-modal semantic alignment, they often provide limited interpretability and may not effectively capture domain-specific contextual information. In environmental event reports, images frequently encode contextual cues that implicitly embed humans' understanding of the most representative concepts that require semantic interpretation rather than direct visual similarity. Recent work in multimodal recommendation and document retrieval suggests that representing visual information through explicit attribute-level or cue-based representations can improve retrieval effectiveness and interpretability \cite{wu2022mm,zhang2024can}. In geospatial domains, spatially meaningful visual cues such as land-use patterns, object distributions, or urban structures have also been shown to provide stronger signals for geographic inference than generic visual features \cite{wu2025advancing}. However, existing multimodal retrieval approaches rarely extract structured geographic context from imagery, limiting their ability to support geographic reasoning in retrieval systems.

\paragraph{Neural Re-ranking.}
Neural re-ranking has become a standard technique for improving retrieval accuracy by applying expressive relevance models to a candidate set produced by an initial retrieval stage. Cross-encoder models represent a widely used architecture in which queries and candidate documents are jointly encoded using transformer-based models to predict relevance scores \cite{reimers2019sentence}. Such models capture rich token-level interactions between queries and documents and are commonly applied in semantic search pipelines. Late interaction models show another direction. For example, ColBERT computes contextualized token embeddings separately for queries and documents and performs fine-grained similarity matching during ranking \cite{khattab2020colbert}. More recently, LLMs have been explored as ranking agents that evaluate candidate relevance using prompting-based reasoning. Approaches such as RankGPT treat ranking as a generative reasoning task in which LLMs score or order candidate documents based on query instructions and contextual evidence \cite{sun2023chatgpt}. Although these methods achieve strong semantic matching performance, they generally focus on textual relevance and rarely incorporate complex context reasoning, such as explicit spatial or temporal context. As a result, semantically similar documents may be ranked highly even when they occur in geographically unrelated contexts or time periods.
\paragraph{Spatiotemporal Event Retrieval.}
Environmental event retrieval requires jointly modeling semantic similarity and spatiotemporal relationships between events. Several studies have proposed integrating geographic distance and temporal proximity into ranking models as additional relevance signals \cite{martins2010learning}. An example is Geo-Time Re-ranking (GT-R) model that computes spatial and temporal offsets between events and incorporate them into ranking functions or apply boosting strategies when candidate events fall within predefined spatial or temporal thresholds \cite{tian2025advancing}. Although these methods demonstrate that spatiotemporal signals improve event retrieval performance, they commonly rely on fixed thresholds that remain constant across queries. In practice, environmental phenomena exhibit substantial variation in spatial and temporal scales. For example, localized ecological disturbances may occur within a small geographic region, whereas regional climate anomaly events may span hundreds of kilometers and persist over extended periods. Fixed thresholds therefore risk excluding relevant events or including unrelated ones when applied across heterogeneous environmental contexts. This limitation highlights the need for retrieval models capable of dynamically adapting spatial and temporal relevance to the characteristics of individual events.

% \paragraph{Summary and Research Gap.}
The literature above demonstrates important advances in geographic information retrieval, multimodal representation learning, and neural re-ranking. Nevertheless, several challenges remain when applying these methods to environmental event retrieval. Existing GIR systems primarily rely on textual signals and static spatial features, limiting their ability to capture the multimodal nature of event reports. Multimodal retrieval models typically employ joint embeddings that provide limited interpretability and rarely extract structured geographic context from images. Meanwhile, neural re-ranking approaches focus primarily on semantic similarity and generally overlook spatial and temporal relationships between events. Finally, existing spatiotemporal retrieval models frequently rely on fixed geographic or temporal thresholds that cannot account for the scale-dependent nature of environmental phenomena. To address these limitations, this work proposes a unified framework that integrates multimodal semantic understanding with adaptive spatiotemporal reasoning. 
Specifically, our approach extracts structured visual cues from event imagery using a VLM and leverages LLM reasoning to infer query-specific spatial and temporal scales for ranking. By combining multimodal retrieval with scale-aware geographic reasoning, the proposed framework advances spatial-temporal event search and recommendation within the context of geographic information retrieval.

\begin{figure*}[t]
    \centering
    \includegraphics[width=2\columnwidth]{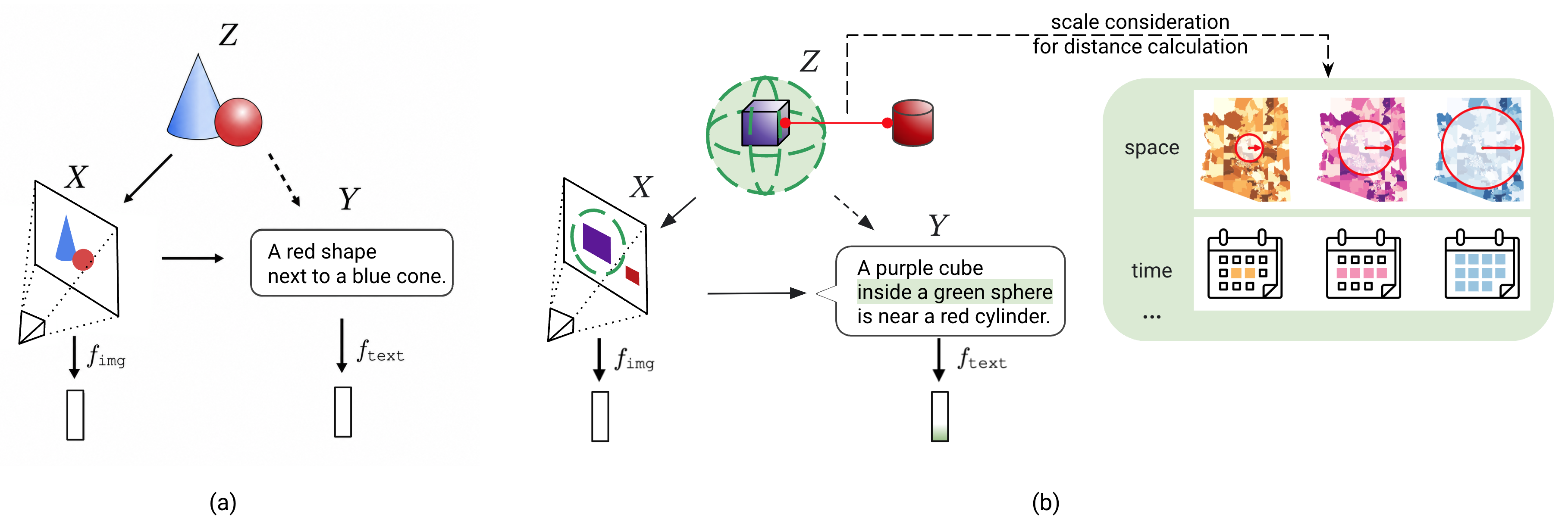} % Replace with your image path
    \caption{Motivation for multimodal context and scale consideration. (a) Platonic representation (figure adopted from \cite{huh2024position}. Original figure licensed under CC BY 4.0.); (b) context-aware and scale-considered Platonic representation (figure of our proposed approach). Image $X$ and text $Y$ are projections of an underlying reality ($Z$).}
    \label{fig:figure1}
\end{figure*}

\section{THEORETICAL FOUNDATION AND CONCEPTUAL DESIGN}

We ground our framework in a theoretical synthesis of multimodal representation and geographic scale-dependency to address the mentioned gaps.

\subsection{Operationalizing Multimodal Convergence}

The conceptual design of our framework is grounded in the Platonic Representation Hypothesis, which posits that representations from different modalities (e.g., image $X$ and text $Y$ in Figure~\ref{fig:figure1}a) are projections of a shared underlying reality ($Z$) and naturally converge toward a unified ideal representation \cite{huh2024position}. We extend this hypothesis to the GIR domain by arguing that for environmental events, this ideal reality $Z$ is inherently multidimensional, encompassing semantic, spatial, and temporal truths that cannot be fully captured by a single modality alone. We operationalize this convergence through the proposed CAMERA algorithm (details in 4.3), which treats VLMs not merely as extractors, but as translators that map visual projections ($X$) into semantically grounded textual cues. By fusing these visual cues with narrative text ($Y$), we reconstruct a more complete approximation of the event’s Platonic representation ($Z$), enabling the retrieval of related events that share the same underlying geographic and thematic context even when explicit keyword matches are absent.

\subsection{Scale-Awareness as a Functional Requirement}
The Platonic hypothesis provides the mechanism for multimodal fusion, and the necessity of scale arises from the core geographic principle of spatial heterogeneity, which dictates that the significance of proximity varies across different environments and phenomena \cite{augusto2022contexts,fotheringham2009geographically,oshan2022scoping}. As discussed in related works, traditional GIR models have the limitation of using static, global, empirical, or manually-defined or rigid thresholds that ignore the multiscale complexity of environmental processes. We propose that an ideal representation of an event should include its unique geographic and temporal scale. This theoretical requirement is operationalized in our ASTRA algorithm (details in 4.4), which addresses this requirement by leveraging LLM-based reasoning to infer query-specific relevance scales. By dynamically adjusting these thresholds based on the event’s specific nature, such as the localized impact of a flash flood versus the regional reach of a drought event, this can make the similarity assessment more flexible and better aligned with the physical reality of the geographic phenomenon.

\section{METHODOLOGY}

\subsection{Preliminaries and Problem Statement}

In this section, we introduce fundamental concepts and definitions that are used throughout the proposed framework for geospatial event recommendation. Key notations are summarized in Table~\ref{tab:notation} for reference.

\begin{table*}[htbp]
\centering
\caption{Summary of key notations used in the CAMERA and ASTRA.}
\label{tab:notation}
\renewcommand{\arraystretch}{1.2}
\begin{tabular}{p{3.5cm} p{10.5cm}}
\toprule
\textbf{Notation} & \textbf{Description} \\
\midrule
$q, C$ & Query event and the set of candidate event documents \\
$\mathcal{L}_1, \mathcal{L}_2$ & Top-$k$ event lists resulting from Stage 1 (Retrieval) and Stage 2 (Re-ranking) \\
$e.\{\mathrm{attr}\}$ & Event attributes: \texttt{title}, \texttt{sum} (summary), \texttt{cat} (category), \texttt{time}, \texttt{loc}, \texttt{img} \\
$v_{\{\mathrm{cat}, \mathrm{loc}, \mathrm{time}\}}$ & Visual cues inferred from event images via VLM \\
$r_e \in \mathbb{R}$ & Dense semantic embedding of event $e$ \\
\midrule
\multicolumn{2}{l}{\textit{ASTRA Spatiotemporal Features}} \\
$\Delta_{geo}, \Delta_{time}$ & Raw physical distances: Haversine distance and absolute day-of-year difference \\
$\mathbb{K}(e)$ & Categorical climate class \\
% $\delta_{clim}$ & Binary climate zone indicator \\
$\delta_{clim}$ & Binary climate zone indicator \\
\midrule
\multicolumn{2}{l}{\textit{Adaptive Scale Parameters and Kernels}} \\
$\sigma_q, \tau_q$ & LLM-inferred spatial and temporal bandwidths (resolution scales) for query $q$ \\
$\kappa_{geo}, \kappa_{time}$ & Adaptive similarity kernels for spatial and temporal domains \\
\midrule
\multicolumn{2}{l}{\textit{Learning-to-Rank Components}} \\
$f_{semantic}, f_{cat}$ & Semantic rank prior and Jaccard-based categorical similarity features \\
$M_{LTR}$ & The Learning-to-Rank model\\
$\mathbf{x}_j \in \mathbb{R}^7$ & Compiled feature vector for candidate $c_j$ ($c_j \in C$) input to $M_{LTR}$ \\
$S_j$ & The resulting final relevance score for $c_j$  \\
\bottomrule
\end{tabular}
\end{table*}

\textbf{Event Definition.} 
An event $e$ is defined as a discrete geospatial phenomenon documented through multimodal reporting. Each event $e \in \mathcal{E}$ is represented by a tuple of structured and unstructured attributes: $e = \langle title, sum, cat, time, loc, img \rangle$. Here, $title$ and $sum$ provide the linguistic context, $cat$ denotes thematic category tags, $time$ and $loc$ provide the spatiotemporal anchors, and $img$ is a representative visual record of the occurrence.

\textbf{Query and Candidate Events.}
Given a query event $q \in \mathcal{E}$, the retrieval task seeks to identify a set of candidate events $\mathcal{C} \subset \mathcal{E} \setminus \{q\}$ that exhibit the highest degree of contextual similarity. In this work, the candidates are initially filtered through a high-recall retrieval stage (CAMERA) to form the subset $\mathcal{L}_1$, which is subsequently refined into the final recommendation list $\mathcal{L}_2$ through the ASTRA re-ranking framework.

\textbf{Multimodal Event Representation.}
To support efficient similarity-based search, each event $e$ is projected into a $d$-dimensional semantic space, resulting in an embedding $\mathbf{r}_e \in \mathbb{R}$. Distinct from traditional multimodal fusion that concatenates latent feature vectors, our approach utilizes a \textit{multimodal-informed} text representation. This is achieved by extracting semantically grounded visual cues $v$ from $e.img$ and concatenating them with textual attributes before encoding. This ensures that the dense representation $\mathbf{r}_e$ captures visual context and remaining compatible with standard dense retrieval backends.

\textbf{Problem Statement: Scale-Aware Event Recommendation.}
The objective is to produce a ranked list $\mathcal{L}$ of top-$k$ events that are most relevant to a query $q$. We define relevance as a non-linear function of multimodal semantic similarity and spatiotemporal proximity. Crucially, we treat this as a \textit{scale-aware} problem: the relevance of a candidate $c_j$ is not only dependent on its distance from $q$, but also on the semantic extent $(\sigma_q, \tau_q)$ of the query event itself. The goal is thus to learn a ranking function $f(q, c_j)$ that dynamically weights these factors to maximize retrieval precision across heterogeneous event types.

\subsection{Method Overview}

\begin{figure*}[htbp]
    \centering
    \includegraphics[width=2\columnwidth]{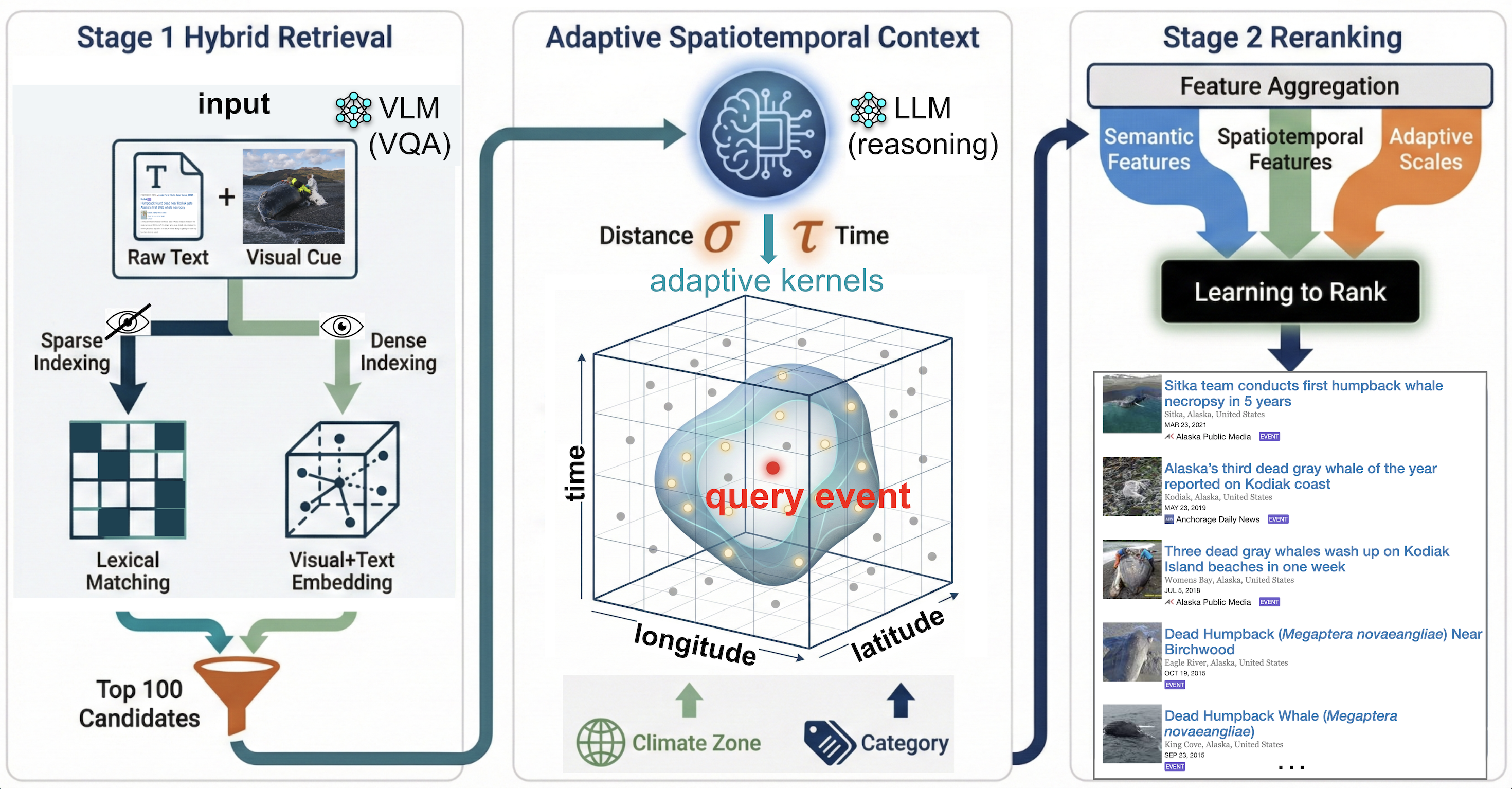} % Replace with your image path
    \caption{Proposed multimodal and multi-scale event retrieval and re-ranking framework. 
    % (a) The use case for the environmental event recommendation. The query event is shown on the left, and the recommended similar events are shown on the right. (b) The proposed multimodal and multiscale framework for recommending similar spatiotemporal events.
    }
    \label{fig:figure2-r1}
\end{figure*}

The proposed framework employs a two-stage architecture (Figure~\ref{fig:figure2-r1}) designed to bridge the gap between multimodal semantics and geographic scale-dependency in event recommendation. During the first stage, \textbf{CAMERA} (Context-Aware Multimodal Event Retrieval Algorithm) generates a high-recall candidate list by encoding textual metadata alongside semantically grounded visual cues through hybrid retrieval. This is followed by \textbf{ASTRA} (Adaptive Spatio-Temporal Re-ranking Algorithm), a supervised Learning-to-Rank stage that refines the selection by modeling non-linear interactions between thematic alignment and physical constraints. ASTRA's core novelty lies in its use of LLM-based reasoning to infer query-specific spatial and temporal bandwidths, which parameterize adaptive kernels to transform raw spatiotemporal distances into context-sensitive relevance scores. Both CAMERA and ASTRA are designed as pipeline components that integrate foundation models (VLMs and LLMs) as modular semantic and reasoning engines.

\subsection{Stage 1: Context-Aware Multimodal Event Retrieval}

\subsubsection{Visual Cues}

The core idea of CAMERA is to construct event representations that capture semantic, spatial, and temporal context from multimodal inputs, through concise and interpretable cues rather than raw visual features, and remaining compatible with standard information retrieval pipelines. 
% Rather than directly fusing latent visual and textual embeddings, 
CAMERA represents visual information through semantically meaningful textual cues that can be integrated into text-based retrieval models.
We define a \emph{cue} as a short textual description corresponding to an important aspect of an event, such as its category or geographic context. As we reviewed in the related works, prior work has shown that incorporating visual cues can enhance representation quality for information retrieval tasks. By expressing visual information in textual form, CAMERA prioritizes interpretability and preserves compatibility with dense, sparse, and ensemble retrieval backends.

\begin{figure*}[htbp]
    \centering
    \includegraphics[width=2\columnwidth]{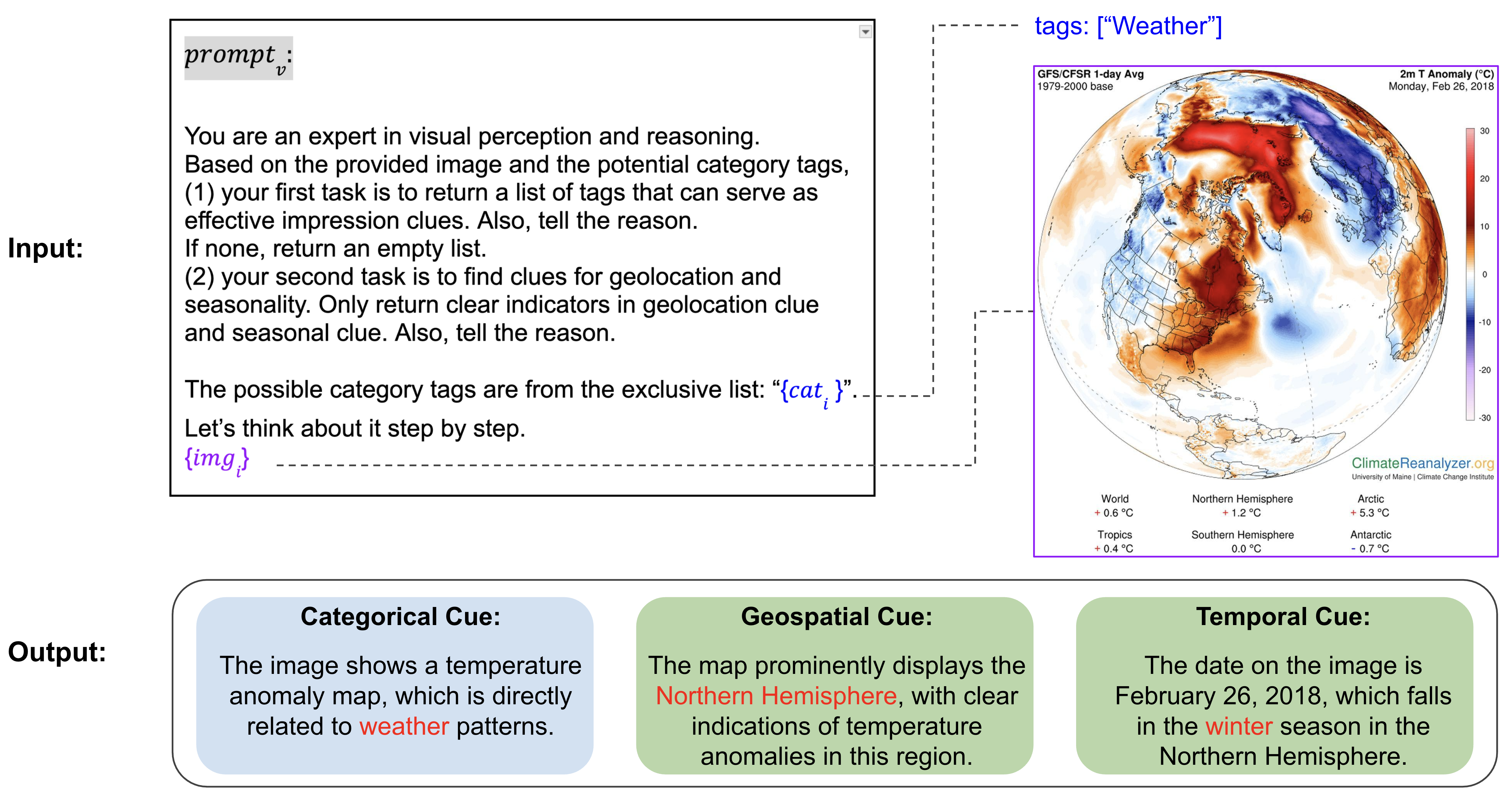} % Replace with your image path
    \caption{Extracting visual cues, including content and context from aspects of event category, location, and time using a VLM. Input includes a prompt, a list of category tags and a thumbnail image of an event. The example is an event of temperature anomalies.}
    \label{fig:figure3}
\end{figure*}

VLM is prompted to infer visual cues from the event thumbnail image related to event category, approximate geographic context, and temporal or seasonal characteristics.  
The prompt specifies the designated expertise for visual extraction by role-playing, provides the event’s category tags as contextual input, and explicitly specifies the cue extraction tasks for each aspect.
It also uses zero-shot chain-of-thought technique to improve extraction performance by outputting the reasoning process through intermediate steps.
Figure~\ref{fig:figure3} illustrates the prompt and an example output for an event reporting above-freezing temperatures at the North Pole, where the input consists of a category tag set (["\textit{Weather}"]), an event image showing the global spatial distribution of temperature anomalies, and a structured prompt requesting extraction. In response, the VLM produces short textual descriptions for each aspect, such as inferred geographic context and seasonality, which serve as visual cues ($v_{cat}$, $v_{loc}$, $v_{time}$).

\begin{figure*}[htbp]
    \centering
    \includegraphics[width=2\columnwidth]{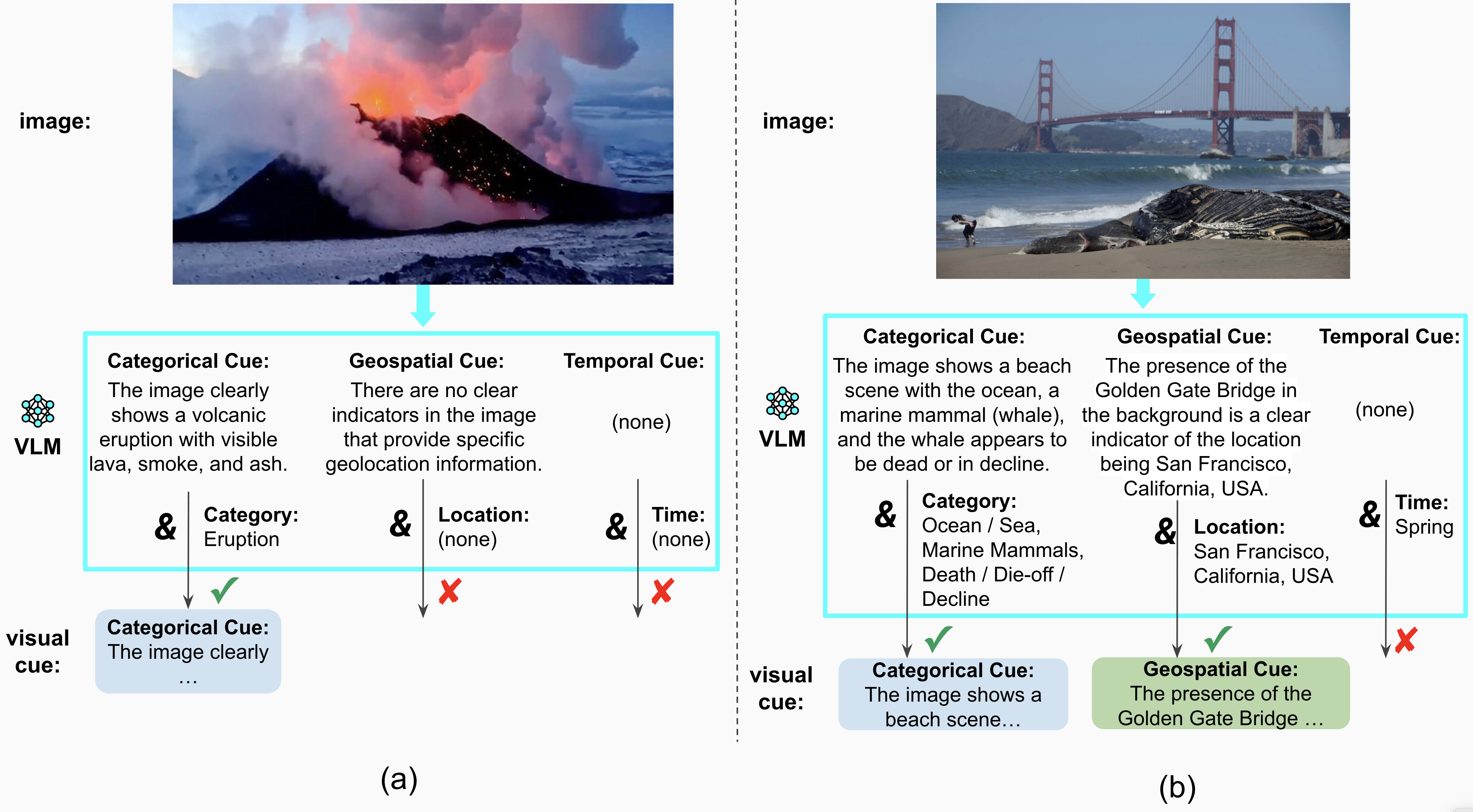} % Replace with your image path
    \caption{
    % Verification process of visual cue extraction. (a) a volcanic eruption; (b) a dead whale in San Francisco
    Automated verification of visual cue extraction via logical intersection. Visual cues are retained only when the VLM-extracted reasoning aligns with inferred categorical, geospatial, or temporal cues (through the AND operation). (a) a volcanic eruption event; (b) an event about the 5th dead whale found in less than month near San Francisco Bay. 
    }
    \label{fig:figure-reflection}
\end{figure*}

To improve robustness, CAMERA designs an automated verification step to guardrail whether each extracted cue is both supported by the image and relevant to the target aspect (category/location/time), related to Algorithm~\ref{alg:retrieval_camera} line 3. 
% a goal-oriented reflection \cite{zeng2024perceive} step
% This checks consistency between the cue and visual evidence, and rejects speculative or weakly supported outputs when the evidence is insufficient.
This is implemented as a logical AND operation between the VLM’s reasoning and the predefined event tags, effectively filtering out speculative outputs without human intervention.
% That means, either extracted aspects are empty (location, time in Figure~\ref{fig:figure-reflection}a), or
% reasoning process is mull (time in Figure~\ref{fig:figure-reflection}b),
% leads to an empty visual cue result for that aspect.
As shown in Figure~\ref{fig:figure-reflection}, if the visual evidence is insufficient (e.g., missing location cues in a volcanic eruption) or the reasoning process is null, the system rejects the output to prevent hallucinations.
This verification process reduces irrelevant or incorrect extractions and yields a more stable set of cues for downstream retrieval. From a representation perspective, the cue-based strategy constitutes a structured multimodal fusion \cite{liang2024foundations} that transforms raw visual input into disentangled, aspect-specific textual descriptors that can be embedded and indexed efficiently.

\subsubsection{Retrieval}

Extracted visual cues are concatenated with textual event descriptions and spatiotemporal metadata to form a unified textual input. Semantic similarity between the query event and candidate events is computed using cosine similarity score over dense embeddings. In parallel, a sparse lexical index based on BM25 is constructed to support keyword-based matching. 
Dense and sparse retrieval scores are combined using an ensemble retriever to generate an initial ranked candidate list, from which the top-$k_1$ events are selected for subsequent re-ranking. 
This approach allows the retrieval to benefit from both semantic understanding of neural embeddings and maintaining and keyword matching capabilities of sparse indexing.

Algorithm~\ref{alg:retrieval_camera} describes the retrieval stage (\textit{Stage~1}) of the proposed framework. Given an event $e_i$, CAMERA constructs a unified textual representation by integrating structured textual attributes with visual cues inferred from the event’s representative image. Textual information $t_{e}$ is constructed by concatenating key attributes and metadata, including $e.title$, $e.sum$, $e.cat$, $e.time$, and $e.loc$.
Then, it concatenates this structured textual information with visual cues. This enriched multimodal-informed sequence is projected into a high-dimensional vector space using a pre-trained text embedding model to generate a dense semantic embedding $r_e$.

To ensure both semantic depth and lexical precision, CAMERA employs an ensemble retrieval strategy. For each candidate event $c_j$ in the collection $C$, the framework computes a pairwise ($q$ and $c_j$) dense similarity score $s^{dense}_j$ based on the cosine proximity of the visual-informed embeddings ($r_q$ and $r_{c_j}$),
and a sparse lexical score $s^{sparse}_j$ using the BM25 algorithm applied directly to the concatenated textual attributes ($t_q$ and $t_{c_j}$). Because these scores reside in different numerical scales, we apply Reciprocal Rank Fusion (RRF) \cite{cormack2009reciprocal} to fuse the resulting rankings robustly. 
Finally, the top-$k_1$ ($k_1$ = 100) events are selected based on their fused ranks to form the candidate set for subsequent processing.

\begin{algorithm}[htbp]
\caption{Context-Aware Multimodal Event Retrieval Algorithm (\textbf{CAMERA})}
\label{alg:retrieval_camera}
\SetKwInput{KwInput}{Input}
\SetKwInput{KwOutput}{Output}
\SetKwFunction{EventEmbedFunc}{EventEmbed}
\SetKwFunction{SpecificWordEmbed}{TextEmbed}
\SetKwFunction{SpecificVLM}{VLM}
\SetKwFunction{SemanticSearch}{SemanticSearch}
\SetKwFunction{BM}{BM25} 
\SetKwFunction{RRF}{RRF} 

\KwInput{
$q$: Query event;\\
$C = \{c_1, \dots, c_m\}$: Candidate events;\\
$prompt_v$: Visual cue extraction prompt;\\
$k_1$: Retrieval count.
}
\KwOutput{$\mathcal{L}_1$: list of retrieved events.}

\SetKwProg{Fn}{Function}{:}{}
\Fn{\EventEmbedFunc{$e$, $prompt_v$}}{
    $t_{e} \gets \oplus(e.title, e.sum, e.cat, e.time, e.loc)$ \tcp*[r]{Textual attributes and metadata}
    $(v_{cat}, v_{loc}, v_{time}) \gets \SpecificVLM(e.img, e.cat, prompt_v)$ \tcp*[r]{Visual cues}
    $t_{multimodal} \gets \oplus(t_{e}, v_{cat}, v_{loc}, v_{time})$ \tcp*[r]{Concatenate textual and visual info}
    $r_e \gets \SpecificWordEmbed(t_{multimodal})$ \tcp*[r]{Embed to space $\mathbb{R}$ }
    \Return $(t_e, r_e)$ 
}

$t_q, r_q \gets \EventEmbedFunc(q, prompt_v)$\; 

\tcp{Compute similarity scores}
\ForEach{$c_j \in C$}{
    $(t_{c_j}, r_{c_j}) \gets \EventEmbedFunc(c_j, prompt_v)$\;
    $s^{dense}_j \gets \SemanticSearch(r_q, r_{c_j})$ \tcp*[r]{Dense retrieval on multimodal rep.}
    $s^{sparse}_j \gets \BM(t_q, t_{c_j})$ \tcp*[r]{Sparse retrieval on text}
}

\tcp{Generate rankings based on raw scores}
$\mathcal{R}_{dense}, \mathcal{R}_{sparse} \gets \text{Rank } C \text{ by } s^{dense}, s^{sparse}$\;

\ForEach{$c_j \in C$}{
    $s_j \gets \RRF(\text{rank}(c_j, \mathcal{R}_{dense}), \text{rank}(c_j, \mathcal{R}_{sparse}))$ \tcp*[r]{Ensemble: rank fusion}
}

$\mathcal{L}_1 \gets \text{Top-}k_1\text{ events ranked by } s_j$\;
\Return $\mathcal{L}_1$
\end{algorithm}

\subsection{Stage 2: Adaptive Spatio-Temporal Re-ranking}
\label{sec:astra}
To refine the candidate set $\mathcal{L}_1$ generated by the initial retrieval stage, we propose \textbf{ASTRA} (Adaptive Spatio-Temporal Re-ranking Algorithm). Although the CAMERA stage optimizes for high recall through multimodal embeddings, it primarily relies on linear fusion and vector proximity, which may not fully capture the complex, non-linear dependencies between event semantics and their spatiotemporal context. Another point is that spatial and temporal relevance are inherently scale-dependent. The appropriate distance or time window for determining event similarity can vary substantially across event types, geographic extents, and temporal dynamics. Fixed relevance thresholds \cite{tian2025advancing} therefore risk either over-filtering or over-including candidates in heterogeneous environmental settings.
ASTRA addresses this limitation by transitioning to a supervised Learning-to-Rank (LTR) framework, LambdaMART \cite{burges2010ranknet} that treats re-ranking as a scale-aware optimization problem. 
It jointly considers semantic relevance and spatiotemporal proximity in a context-sensitive manner.

\begin{figure*}[htbp]
    \centering
    \includegraphics[width=2\columnwidth]{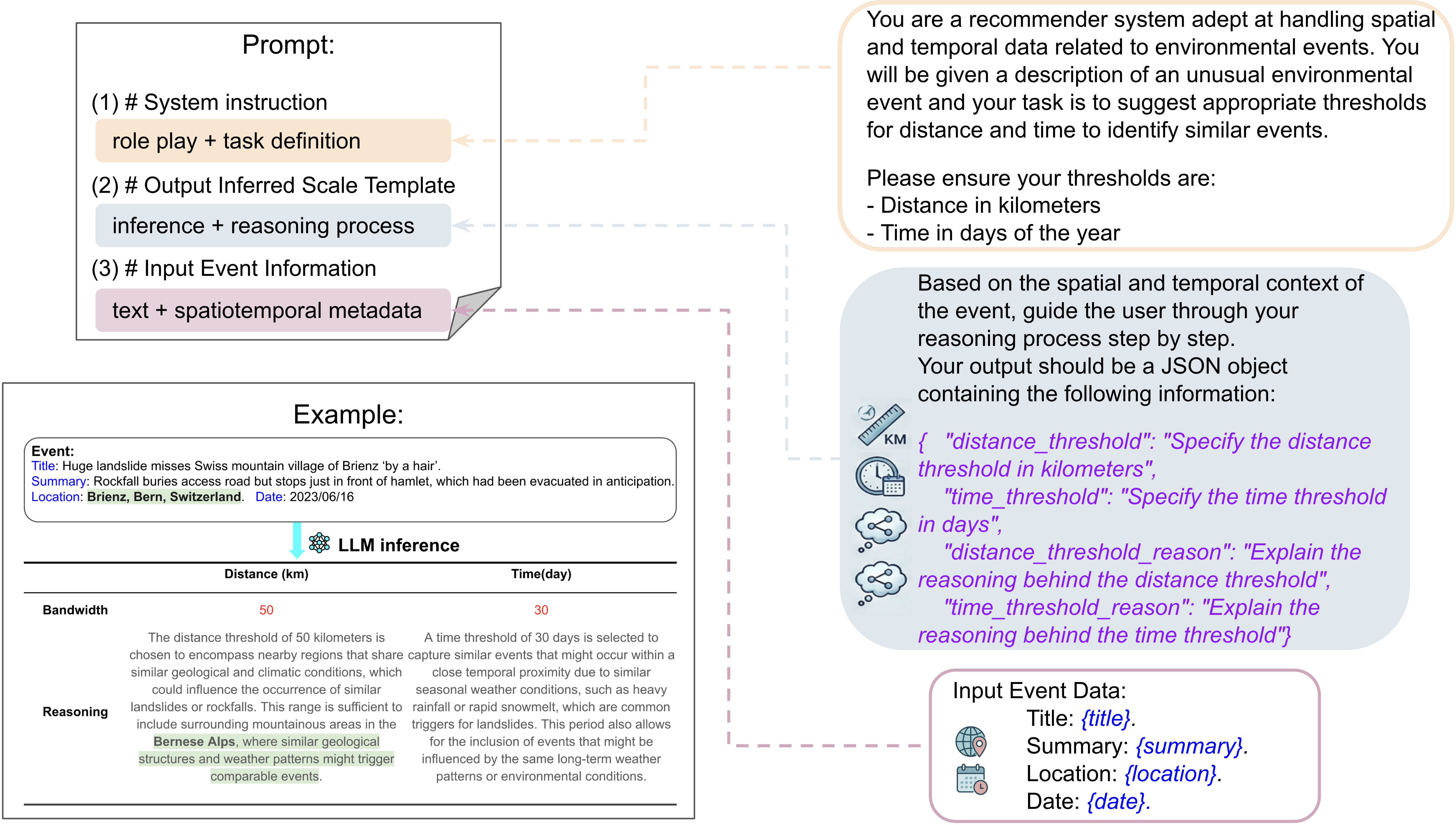} % Replace with your image path
    \caption{Prompt for ASTRA: adaptive scale inference by LLM}
    \label{fig:figure-ASTRA_prompt}
\end{figure*}

The novelty of ASTRA lies in the idea of moving from static distance thresholds to query-specific adaptive kernels. This bridges a critical gap between the geography and AI communities. In geographic information science, the scale-dependency of event relevance is a fundamental challenge. The semantically appropriate search radius, or bandwidth, varies across event types. For instance, a localized landslide affects a kilometer scale, whereas a regional wildfire or a tsunami needs a vastly broader spatiotemporal scale of attention. ASTRA operationalizes this geographic intuition by utilizing the reasoning capabilities of LLMs to perform zero-shot inference of optimal geospatial and temporal bandwidths ($\sigma_q, \tau_q$) for each query. 
This approach allows the re-ranking model to dynamically modulate its inductive bias, transforming raw spatiotemporal distances into adaptive similarity kernels that reflect the true semantic extent of the query event. ASTRA operates fully automatically at inference time. 
The prompt used for adaptive threshold inference (Figure~\ref{fig:figure-ASTRA_prompt}) and includes structured event attributes such as location and timestamp, with a fixed output format and explicit units. All LLM inference is performed using deterministic decoding settings to ensure reproducibility across runs.

\paragraph{Feature Set: LLM-Based Adaptive Scaling} 
The primary novelty of ASTRA is the dynamic inference of event-specific resolution parameters. Unlike static re-ranking models, ASTRA utilizes LLM reasoning to infer a spatial bandwidth $\sigma_q$ and a temporal bandwidth $\tau_q$ for each query. These parameters represent the semantic extent of the event. We operationalize these scales by passing the raw distances ($\Delta_{geo}, \Delta_{time}$) through  Gaussian radial basis (GRB) function kernels (Algorithm~\ref{alg:reranking_astra}, line 9 and 10):
\begin{equation}
    \kappa_{geo} = \exp\left( -\frac{\Delta_{geo}^2}{2\sigma_q^2} \right), \quad \kappa_{time} = \exp\left( -\frac{\Delta_{time}^2}{2\tau_q^2} \right)
\end{equation}
By transforming raw distances ($\Delta$) into normalized adaptive kernels ($\kappa \in [0, 1]$), ASTRA provides the re-ranking model with 
% an inductive bias 
contextual constraint sensitive to the query's inherent scale.

As detailed in Algorithm~\ref{alg:reranking_astra}, ASTRA represents each query-candidate pair $(q, c_j)$ as a feature vector $\mathbf{x}_j$ composed of three distinct functional dimensions of event similarity. Below are the feature taxonomy and rationalization.

\begin{algorithm}[htbp]
\caption{Adaptive Spatio-Temporal Re-ranking Algorithm (\textbf{ASTRA})}
\label{alg:reranking_astra}
\SetKwInput{KwInput}{Input}
\SetKwInput{KwOutput}{Output}
\SetKwFunction{LLM}{LLM\_Inference}
\SetKwFunction{Haversine}{Haversine}
\SetKwFunction{DOY}{DOY}
\SetKwFunction{Jaccard}{Jaccard}
\SetKwFunction{LTR}{LTR\_Model}

\KwInput{
$q$: Query event;\\
$\mathcal{L}_1$: Top-$k_1$ candidate events from Stage 1;\\
$prompt_s$: Adaptive scale inference prompt;\\
$M_{LTR}$: Trained Learning-to-Rank model.
}
\KwOutput{$\mathcal{L}_2$: Final top-10 re-ranked events.}

\tcp{Step 1: Infer Event-Specific Bandwidths via LLM}
$\sigma_{q}, \tau_{q} \gets \LLM(q, prompt_s)$ \tcp*[r]{Adaptive geospatial and temporal scales}

\tcp{Step 2: Feature Engineering for Candidates}
$\mathcal{F} \gets \emptyset$\;
\ForEach{$c_j \in \mathcal{L}_1$}{
    % \tcp{Group 1: Semantic \& Categorical}
    \tcp{\colorbox{MySemanticBlue}{\textcolor{black}{\textbf{Feature Set \textcircled{1}: Multimodal Semantic Relevance}}}}
    % $f_{rank} \gets \text{rank}(c_j, \mathcal{L}_1)$ \tcp*[r]{Stage 1 rank as a prior}
    $f_{semantic} \gets \text{rank}(c_j, \mathcal{L}_1)$ \;
    $f_{cat} \gets \Jaccard(q.cat, c_j.cat)$\;
    
    % \tcp{Group 2: Raw Context}
    \tcp{\colorbox{MySemanticBlue}{\textcolor{black}{\textbf{Feature Set \textcircled{2}: Physical Spatiotemporal Constraints}}}} 
    $\Delta_{geo} \gets \Haversine(q.loc, c_j.loc)$\;
    $\Delta_{time} \gets |\DOY(q.time) - \DOY(c_j.time)|$\;
    $\delta_{clim} \gets \mathbb{I}(\mathbb{Z}(q) = \mathbb{Z}(c_j))$\;
    
    % \tcp{Group 3: Adaptive Context}
    % \tcp{\colorbox{MySemanticBlue}{\textcolor{black}{\textbf{Group \textcircled{3}: LLM-Based Adaptive Scaling}}}} 
    % $\kappa_{geo} \gets \exp(-\Delta_{geo}^2 / (2\sigma_{q}^2))$\;
    % $\kappa_{time} \gets \exp(-\Delta_{time}^2 / (2\tau_{q}^2))$\;
    \tcp{\colorbox{MySemanticBlue}{\textcolor{black}{\textbf{Feature Set \textcircled{3}: LLM-Based Adaptive Scaling}}}} 
    $\kappa_{geo} \gets \text{RadialBasis}(\Delta_{geo}, \sigma_{q})$ \tcp*[r]{Adaptive geospatial relevance}
    $\kappa_{time} \gets \text{RadialBasis}(\Delta_{time}, \tau_{q})$ \tcp*[r]{Adaptive temporal relevance}
    
    \tcp{Assemble Features}
    $x_j \gets [f_{semantic}, f_{cat}, \Delta_{geo}, \Delta_{time}, \delta_{clim}, \kappa_{geo}, \kappa_{time}]$\;
    $\mathcal{F} \gets \mathcal{F} \cup \{(c_j, x_j)\}$
}

\tcp{Step 3: Supervised Scoring}
\ForEach{$(c_j, x_j) \in \mathcal{F}$}{
    $score_j \gets \LTR(x_j, M_{LTR})$\;
}

$\mathcal{L}_2 \gets \text{Top-10 events ranked by } score_j$\;
\Return $\mathcal{L}_2$
\end{algorithm}

\paragraph{Feature Set: Multimodal Semantic Relevance} 
This set captures thematic alignment and retrieval confidence. We include the initial retrieval rank ($f_{semantic}$) as a hybrid semantic prior. To capture lexical overlap in event metadata, categorical similarity $f_{cat}$ is measured via the Jaccard coefficient of the query and candidate category tags. By integrating $f_{cat}$, ASTRA ensures that candidates sharing specific domain-level attributes are prioritized, complementing the broader semantic proximity captured by the semantic prior.
\begin{equation}
    f_{cat}(q, c_j) = \frac{|q.cat \cap c_j.cat|}{|q.cat \cup c_j.cat|}
\end{equation}

\paragraph{Feature Set: Physical Spatiotemporal Constraints} 
To anchor the re-ranking process in physical reality, we extract raw spatiotemporal distances that provide absolute grounding for the LTR model. Geographic distance ($\Delta_{geo}$) is computed using the Haversine formula to account for the Earth's curvature:

\begin{equation}
\begin{split}
\Delta_{\text{geo}}(q, c_j)
&= 2R \arcsin\left(
\sqrt{
\sin^2\left(\frac{\phi_{c_j} - \phi_q}{2}\right)
+ \cos(\phi_q)\cos(\phi_{c_j})
}
\right. \\
&\qquad\left.
\times \sin^2\left(\frac{\lambda_{c_j} - \lambda_q}{2}\right)
\right)
\end{split}
\end{equation}

where $(\phi, \lambda)$ denote latitude and longitude in radians, and $R$ is the Earth's mean radius ($\approx 6371$ km). Temporal offset ($\Delta_{time}$) is defined as the absolute difference in days-of-the-year ($\text{DOY}$) between the query and candidate event, where $T$ represents period of the cycle (365 days):
\begin{equation}
\Delta_{\text{time}}(q, c_j) = \min \left( \left| \text{DOY}(q) - \text{DOY}(c_j) \right|, \, T - \left| \text{DOY}(q) - \text{DOY}(c_j) \right| \right)
\end{equation}

Environmental events such as floods separated by about 1,000 km can be closely related when driven by the same large-scale atmospheric circulation and climate zone. In contrast, geographically closer events may be weakly related if they arise from different climatic or hydrological regimes.
To capture ecological context beyond geographical distance, a binary climate match indicator ($\delta_{clim}$) is built to enrich context with environmental regimes. Given that high-resolution climate zone predictions contain null values at coastal boundaries or data-sparse regions, a tiered spatial matching method is used to determine the Köppen-Geiger class $\mathbb{K}(e)$ for every event. Let $\mathcal{R}clim = \{1, 10, 50, 100\}$ represent the set of available raster resolutions (in km) of climate zones. Categorical climate value ($V_res(e)$)  is sampled from the raster at resolution $res \in \mathcal{R}clim$. The final climate class $\mathbb{Z}(e)$ is assigned using a prioritized lookup of the highest available resolution based on event coordinates:
\begin{equation}
    \mathbb{Z}(e) = V_{res^*}(e), \quad \text{where } res^* = \min \{res \in \mathcal{R}clim \mid V_{res}(e) \neq \text{NaN}\}
\end{equation}
This matching method ensures spatial robustness by falling back to coarser global environmental regimes when local-scale data is unavailable. The final feature is then formulated as the equivalence of the query and candidate classes (Algorithm~\ref{alg:reranking_astra}, line 8).

\section{EXPERIMENTS}
\subsection{Data Source}

\begin{figure*}[htbp]
    \centering
    \includegraphics[width=2\columnwidth]{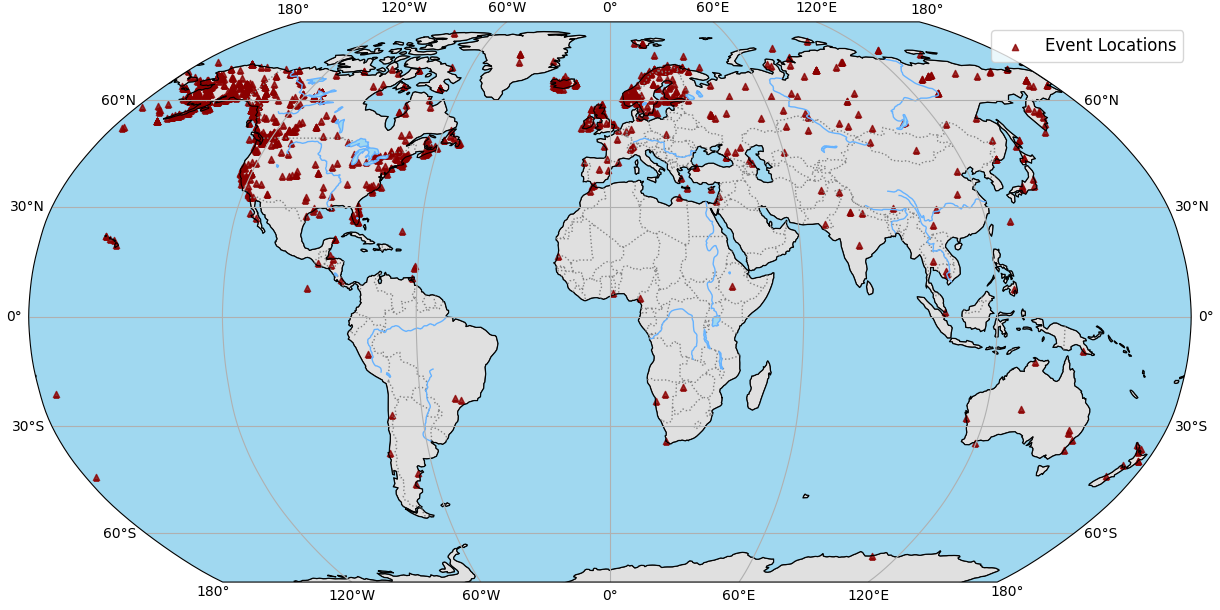} % Replace with your image path
    \caption{Global coverage of event locations in the Local Environmental Observer (LEO) Network}
    \label{fig:figure_LEON_map}
\end{figure*}

\begin{figure*}[htbp]
    \centering
    \includegraphics[width=2\columnwidth]{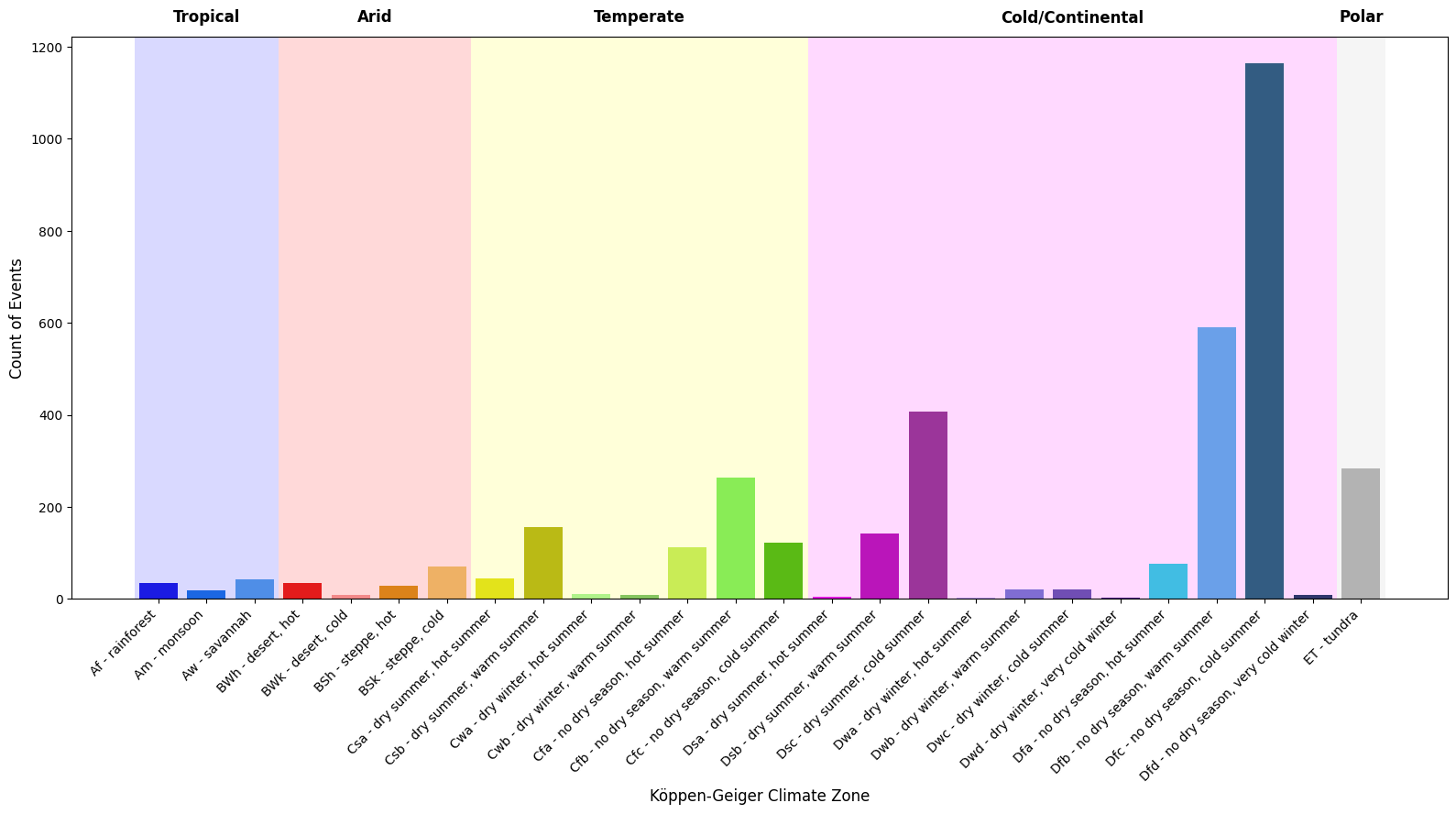} % Replace with your image path
    \caption{Climate zone coverage of events in the dataset}
    \label{fig:figure-climate-zone}
\end{figure*}

Experiments use data from the Local Environmental Observer Network (LEO), an open web platform that shares unusual environmental, climate, and animal events from local news and reports \cite{brubaker2013leo}. 
We chose and built this dataset because it represents a complex, real-world testbed for the challenges of modern environmental monitoring.
The LEO Network collects information on how changing environments impact local communities, offering a collaborative, community-driven perspective that is vital for social and ecological observation \cite{danielson2022monitoring,griffith2018community}.
By engaging a diverse range of observers, including local members, scientists, and policymakers, the platform provides a unique, high-quality repository that expanded from its Arctic origins to cover global events  \cite{mosites2018environmental}.
% The data repository was built for the Arctic regions and later expanded to cover global events. 
The experiments were conducted on all available multilingual records of 4,575 events at the time of writing (Figure~\ref{fig:figure_LEON_map}), which cover 26 climate zones from rain forest to tundra (Figure~\ref{fig:figure-climate-zone}).

Each event record includes a title, a descriptive summary, and category tags, along with comprehensive metadata including place name, geographic coordinates, and date. An advantage of this dataset is that most entries include representative images selected through expert-level curation, providing a gold standard for multimodal analysis.
% Evaluation uses a benchmark of 1,000 query events.
Each valid query event is mapped to relevant events labeled by annotators with domain expertise, ensuring that our performance assessments are grounded in expert-verified geographic and thematic contexts.

\subsection{Dataset Construction and Leakage Prevention}
\label{sec:dataset}

To train and evaluate the re-ranking framework, the raw multimodal event corpus is transformed into a supervised learning format by chronological partitioning, rigorous leakage prevention, and pairwise feature engineering.

% \textbf{Chronological Partitioning:} 
A chronological splitting strategy is used to simulate a real-world recommendation scenario where the system must retrieve historical context for a newly occurring event. As summarized in Table~\ref{tab:dataset_split}, the test set was defined as the 1,000 most recent labeled events in the corpus, spanning May 2020 to October 2023. The remaining events were split into a training set and a validation set at an 80/20 ratio. This ensures a strict temporal ordering where $\text{Train} < \text{Val} < \text{Test}$, maintaining the causal integrity. Here < means ``older than.''

% \textbf{Leakage Prevention:} 
Another challenge in this news-like recommendation is the presence of ground-truth links (e.g., similar events listed in "See Also" on LEON platform) that might inadvertently reveal test-set relationships during the training phase. To ensure the model learns to generalize from spatiotemporal and semantic features rather than memorizing event IDs, a rigorous leakage removal is conducted. Any historical event ID appearing in the ground-truth relevance lists of the Test Set was purged from the training and validation pools. This step removed 1729 potential leakage events, ensuring zero overlap between the testing ground-truth and the training candidates.

% \textbf{Pairwise Feature Engineering and Hard Negative Mining:} 
For each query $q$ within the splits, we constructed a pairwise dataset for supervised learning. Positive samples were drawn from ground-truth relevance links. To challenge the discriminative capacity of the ASTRA re-ranker, we performed \textit{hard negative mining} by selecting the top 50 candidates returned by the Stage 1 CAMERA retrieval phase that were not ground-truth matches, resulting in a final dataset of approximately 80k pairs.

\begin{table*}[htbp]
\centering
\caption{Summary of the dataset split, temporal distribution, and pairwise label distribution for ASTRA re-ranking.}
\label{tab:dataset_split}
\renewcommand{\arraystretch}{1.2}
\begin{tabular}{l c c c c}
\toprule
\textbf{Data Split} & \textbf{Date Range} & \textbf{Queries ($q$)} & \textbf{Positive Pairs <$q$, $c$>} & \textbf{Hard Negative Pairs <$q$, $c$>} \\
\midrule
Training & 1998/05 -- 2019/11 & 1,881 & 1,820 & 23,487 \\
Validation & 2019/11 -- 2023/09 & 471 & 188 & 3,545 \\
Testing & 2020/05 -- 2023/10 & 1,000 & 2,925 & 47,472 \\
\midrule
\textbf{Total} & 1998/05 – 2023/10 & \textbf{3,352} & \textbf{4,933} & \textbf{74,504} \\
\bottomrule
% \multicolumn{5}{p{13cm}}{\footnotesize \textit{Note: 1,729 events were removed during the Leakage Prevention step to ensure zero overlap between the Test Set ground-truth and the Training pool.}}
\end{tabular}
\end{table*}

\subsection{Evaluation Metrics}
\label{sec:evaluation}

The performance evaluation of the proposed two-stage framework employs a suite of standard information retrieval metrics. Given the different objectives of each stage, initial retrieval phase (CAMERA) is evaluated on candidate coverage and the re-ranking phase (ASTRA) is evaluated on fine-grained ordering quality within the top-$k$ results.

% \paragraph{Retrieval Metrics (Stage 1)}
The primary goal of the retrieval phase is to maximize the inclusion of relevant events within the initial candidate shortlist $\mathcal{L}_1$. Metrics include \textbf{Recall@$k$} and \textbf{Hit Rate@$k$} ($k=100$). Recall quantifies the proportion of ground-truth relevant events successfully captured in the top-$k$ results. Hit Rate measures the percentage of queries for which at least one relevant event is retrieved.

% \paragraph{Re-ranking Metrics (Stage 2)}
The re-ranking stage aims to maximize the utility of the final recommendations, so the precise positioning of relevant events is important. Metrics are \textbf{nDCG@$k$}, \textbf{MRR@$k$}, and \textbf{MAP@$k$} at $k \in \{1, 5, 10\}$. MRR (Mean Reciprocal Rank) focuses on the position of the first relevant result, nDCG (Normalized Discounted Cumulative Gain) and MAP (Mean Average Precision) provide a holistic evaluation of the overall ranking order, specifically penalizing models that place relevant events at lower ranks.

% \paragraph{Mathematical Formalization}
Let $Q$ denote the set of all query events. For a given query $q \in Q$, let $R_q$ be the set of ground-truth relevant events in the corpus. We define $\text{rel}(q, i) \in \{0, 1\}$ as an indicator function that is 1 if the event at rank $i$ is relevant to $q$, and 0 otherwise.

The retrieval metrics are defined as follows:
\begin{equation}
    \text{Recall}@k = \frac{1}{|Q|} \sum_{q \in Q} \frac{\sum_{i=1}^k \text{rel}(q, i)}{|R_q|}
\end{equation}
\begin{equation}
    \text{Hit Rate}@k = \frac{1}{|Q|} \sum_{q \in Q} \mathbb{I}\left( \sum_{i=1}^k \text{rel}(q, i) > 0 \right)
\end{equation}

The re-ranking metrics are defined as follows:
\begin{equation}
    \text{nDCG}@k = \frac{1}{|Q|} \sum_{q \in Q} \frac{\text{DCG}_q@k}{\text{IDCG}_q@k}, \quad \text{where } \text{DCG}_q@k = \sum_{i=1}^k \frac{2^{\text{rel}(q, i)} - 1}{\log_2(i+1)}
\end{equation}
\begin{equation}
    \text{MRR}@k = \frac{1}{|Q|} \sum_{q \in Q} \frac{1}{\text{rank}_q}, \quad \text{rank}_q = \min \{i \mid \text{rel}(q, i) = 1\}
\end{equation}
\begin{equation}
    \text{MAP}@k = \frac{1}{|Q|} \sum_{q \in Q} \left( \frac{1}{|R_q|} \sum_{i=1}^k \text{P}_q@i \times \text{rel}(q, i) \right)
\end{equation}
where $\text{P}_q@i = \frac{1}{i} \sum_{j=1}^i \text{rel}(q, j)$ denotes the precision at rank $i$ for query $q$. For all metrics, if no relevant event is found within the top-$k$ results, the corresponding query score is set to 0.

\subsection{Comparing Models and Implementation Details}

During the retrieval stage, our approach employs the GPT-4o model for visual cue extraction in CAMERA due to quality, consistency, and cost-effectiveness consideration, 
with temperature as zero to increase generation consistency and reproducibility. After concatenation, cosine similarity is calculated over bi-encoding (i.e., independently encoding queries and documents into a shared embedding space) on event embeddings generated by OpenAI’s embedding model as a multimodal representation. As a comparison, we use a combination of all textual segments as unimodal input. We compare it against three retrieval paradigms: 

\begin{itemize}
    \item \textbf{Sparse retrieval}: BM25 \cite{robertson2009probabilistic} is an established model for ranking textual content in recommendation tasks. This lexical search method assigns relevance scores to candidate text pieces using a weighting scheme that considers term frequency and document length, balanced by how common the term occurs across the entire collection. BM25 is valued for its simplicity, efficiency, and transparency, making it a strong baseline in many retrieval applications, particularly those that involve explicit keyword queries.
    \item \textbf{Dense Retrieval}: These approaches map queries and candidates into a shared latent space to capture semantic relationships. This method has become a central component in modern recommendation pipelines because it captures deeper contextual and lexical relationships \cite{zhao2024dense}. 
    We evaluate a commercial model as 
    OpenAI embedding model \texttt{text-embedding-3} \cite{neelakantan2022text} and an open-source model \texttt{BGE-m3}\cite{chen2024bge}.
    \item \textbf{Multimodal Retrieval}: 
    To establish a baseline for native vision-language alignment,
    we implement a retrieval baseline using CLIP (\textit{Contrastive Language-Image Pre-training}) \cite{radford2021learning}, the \texttt{clip-vit-large-patch14} architecture.
    We use the pre-trained CLIP model to represent the performance of standard, joint latent space embeddings.
    This configuration serves to evaluate the efficacy of implicit multimodal alignment, which is common in general-purpose foundation models, against the explicit, semantically grounded visual cue extraction in our proposed CAMERA framework.
\end{itemize}

For ASTRA in the re-ranking stage, our approach uses GPT-4 Turbo for adaptive scale generation. This version is used based on an initial quality check comparing with GPT-4o and GPT-4o mini. 
% We leveraged OpenAI’s batch API for faster speed and lower cost for offline computing. 
For re-ranking performance, we compare our model with three categories of semantic-based re-ranking models: 

\begin{itemize}
    \item \textbf{Cross-encoding method}: a cross-encoder typically employs a Transformer-based architecture to jointly process a query and a candidate by feeding them as a single concatenated input, and directly predicts a relevance score. This design allows the model to attend to interactions between all tokens across the pair, enabling a more accurate estimation of semantic relevance. This approach is commonly used during re-ranking in recommender systems. Our study evaluates both open-source and proprietary cross-encoding models.
    \begin{itemize}
        \item \underline{Open-source}: free to use, such as MiniLM model from SentenceTransformer \cite{reimers2019sentence}. It is pre-trained on the Microsoft MARCO dataset constructed from queries and answers logged by a real-world search engine, thus good at semantic search to find passages relevant to the search query. We also compared BGE re-ranking models \cite{chen2024bge, li2023making}.   
        \item \underline{Commercial}: fast, costly, blackboxed. We use re-ranking models from Cohere \cite{cohere_rerank_v3} and Jina\cite{jina_rerank_v2} that use a cross-encoding architecture based on LLMs. It is designed to be used after dense retrieval to surface answers more precisely.
    \end{itemize}
    \item \textbf{Late interaction method}: we include ColBERT (Contextualized Late Interaction over BERT) \cite{khattab2020colbert}, which employs a late interaction mechanism over token-level embeddings to enable fine-grained semantic matching. Nowadays, ColBERT is a popular method for re-ranking in LLM applications.
    \item \textbf{LLM as re-ranker method}: a method that uses LLMs' generative capability to judge text relevance by prompt engineering. It constructs a prompt consisting of relevance instructions, a query event, and candidate events, then prompts an LLM asking for scores or rankings. We employed flagship LLM models of OpenAI to implement this method for both approaches below, including GPT-4o mini and GPT-5.2 models.
    \begin{itemize}
        \item \underline{Pointwise approach}: a naive approach that generates a relevant score for each query-candidate pair \cite{liang2022holistic}. The result list is sorted by generated scores in descending order.
        \item \underline{Listwise approach}: improved relevance judgment by providing a list of candidates rather than one candidate for each relevance comparison. It provides a more holistic view of the candidate pool and generates a re-ranked candidate list. In this category, we employ RankGPT \cite{sun2023chatgpt} that uses LLMs as re-ranker.
    \end{itemize}
    % \item \textbf{Spatiotemporal event re-ranking method}: Geo-Time Re-ranking (GT-R) model explicitly integrates spatiotemporal dimensions with semantic similarity to recommend environmental events \cite{tian2025advancing}. It applies fixed thresholds for spatial and temporal relevance, boosting scores for events within specified distances and time ranges. The thresholds are pre-defined manually.
    \item \textbf{
    {Spatiotemporal event re-ranking method}}: The Geo-Time Re-ranking (GT-R) model \cite{tian2025advancing} 
    % serves as the feature prior 
    provides the foundational logic 
    for our approach. GT-R explicitly integrates spatiotemporal dimensions with semantic similarity using static, manually pre-defined thresholds for spatial and temporal relevance, and then conducts ranking fusion. However, it lacks the flexibility to adapt to varying event scales. In contrast, our proposed ASTRA model extends this logic by constructing dynamic, LLM-inferred adaptive features, allowing for context-specific resolution across diverse event types and contexts.
\end{itemize}

\subsection{Experimental Results and Analysis}

\subsubsection{Retrieval Performance Analysis}
The performance of the Stage 1 retrieval phase is summarized in Table~\ref{tab:retrieval_performance}. Analysis is across retrieval coverage 
% , multimodal synergy, 
and operational efficiency, and the results demonstrate the effectiveness of the proposed CAMERA framework in capturing a high-quality candidate set. Analysis is across retrieval coverage, multimodal synergy, and operational efficiency.

% \paragraph{Retrieval Coverage and Recall}
As shown in Table~\ref{tab:retrieval_performance},  CAMERA serves as a high-performance retriever, achieving the best Recall@100 of 93.4 and a Hit Rate@100 of 98.30. These metrics suggest that the framework successfully establishes a great upper bound of relevant events for the subsequent re-ranking stage, effectively minimizing the risk of early-pipeline information loss.

% \paragraph{Efficacy of Multimodal Representation}
The performance improvement of CAMERA compared with a strong text-only baseline ($R@100=0.922$ by Embed-3-Large) suggests that the explicit extraction of semantically grounded visual cues provides non-redundant contextual information that dense textual embeddings alone cannot capture. Furthermore, the significant underperformance of the native multimodal retrieval baseline ($R@100=0.762$ by CLIP) highlights an insight. Although joint vision-language embeddings are powerful for general tasks, they lack the ability to extract important thematic and spatiotemporal factors, such as "temperature anomaly" related to category tag "weather" (Figure~\ref{fig:figure3}), at the granularity needed to support an effective retrieval process. By leveraging a VLM for structured cue extraction rather than relying on black-box vector alignment, CAMERA bridges the gap between raw visual data and event reasoning, ensuring a robust and comprehensive candidate shortlist.

% \paragraph{Operational Efficiency and Scalability}
Beyond retrieval performance, we analyze the computational efficiency of CAMERA. As shown in Table 3, when visual cues are precomputed for the document corpus, the retrieval stage has a latency of 0.542s, which is consistent with the baseline text-embedding-3-large model used as our dense retrieval. For a new incoming query, the overall latency is 4.207s, reflecting the additional time required for the VLM to perform on-the-fly visual cue extraction. This breakdown shows that although multimodal reasoning introduces an initial inference overhead, the core retrieval remains highly scalable for large-scale environmental monitoring.
Furthermore, the cost of CAMERA (\$4.50 per 1k query) is mainly to pay VLM, dense indexing, and query embedding because lexical indexing by BM25 is free. This combination of high accuracy and low operational overhead makes CAMERA a great retrieval backbone for large-scale, context-aware event recommendation systems.

\begin{table*}[ht]
\centering
\caption{Retrieval Performance Comparison (Top-100)}
\label{tab:retrieval_performance}
\small
\setlength{\tabcolsep}{4pt} 
\begin{tabular}{l c cc cc}
\toprule
\textbf{Method} & \textbf{Modality} & \textbf{Recall@100} & \textbf{Hit Rate@100} & \textbf{Latency} & \textbf{Cost} \\
& & \textbf{(\%)} & \textbf{(\%)} & \textbf{(second/query)} & \textbf{(\$/1k query)} \\
\midrule
BM25\cite{robertson2009probabilistic} & Text & 86.8 & 95.4 & 0.001 & free \\
BGE-m3\cite{chen2024bge}                   & Text & 88.1 & 97.2 & 1.032 & free \\
Embed-3-Large\cite{neelakantan2022text}   & Text & 92.2 & 98.2 & 0.542 & 0.01 \\
\midrule
CLIP\cite{radford2021learning}        & Multimodal & 76.2 & 91.3 & 0.007 & free \\
\textbf{CAMERA (Ours)}                & \textbf{Multimodal} & \textbf{93.4} & \textbf{98.3} & \textbf{4.207*} & \textbf{4.50} \\
\bottomrule
\multicolumn{6}{l}{\footnotesize * Includes 3.665s for VLM-based visual cue extraction and 0.542s for semantic retrieval.} \\
\end{tabular}
\end{table*}

\subsubsection{Re-ranking Performance Analysis}
\label{subsec:reranking_results}

Table~\ref{tab:reranking_grouped} presents a comprehensive performance comparison between our proposed re-ranking model ASTRA and several other re-ranking paradigms. The empirical results reveal several insights into the challenges of spatiotemporal retrieval.

% \textbf{Superiority in Ranking Accuracy.} 
ASTRA achieves the highest performance across most evaluation metrics (e.g., nDCG@10 of 63.8\%, MRR@10 of 72.1\%, and MAP@10 of 51.8\%), significantly outperforming both traditional and LLM-based baselines. 
% Specifically, ASTRA reaches an nDCG@10 of 63.8\% and an MRR@10 of 72.1\%. 
In contrast, many established Cross-Encoder models, either open-source (BGE-reranker, SentenceTransformer) or commercial ones (Jina, Cohere), fail to surpass the performance of the Stage 1 retrieval results. This performance gap suggests that standard semantic encoding is insufficient for capturing the complex, multiscale dependencies inherent in the geographic event relevance.

% \textbf{Comparison with GeoAI and LLM Baselines.} 
Although LLMs used as re-rankers (e.g., RankGPT) show an improvement over Cross-Encoders, they still lag behind our model. For instance, RankGPT achieves an nDCG@10 of 48.0\%, which is around 15.8\% lower than ASTRA. This discrepancy uncovers an important finding for the GeoAI community. General-purpose foundation models have strong reasoning capabilities, but they lack the specialized spatiotemporal grounding required to rank events accurately across varying geographical contexts.

% \textbf{Computational Efficiency and Real-time Viability.} 
Another advantage of ASTRA is its computational efficiency, which matters for a re-ranking algorithm in production. As shown in the \textit{Latency} column of Table~\ref{tab:reranking_grouped}, LLM-based approaches are significantly slower because of inference need, with response times ranging from 14.47 to 23.11 seconds per query. ASTRA’s re-ranking latency is low (1.80 s per query) once the model is trained. 
This speedup over LLM-based methods positions ASTRA as a scalable solution for real-time spatiotemporal applications.

% \textbf{Efficacy of Multiscale Spatiotemporal Awareness.} 
The success of ASTRA can be attributed to its ability to treat geospatial and temporal relevance as dynamic, multiscale factors rather than static constraints. By integrating multimodal representations with these adaptive scales, ASTRA identifies relevant events that are often overlooked by semantic-only models, including late interaction model (ColBERT) and cross-encoders. These results validate our hypothesis that incorporating spatiotemporal impact scales can improve event-centric GeoAI applications.

\begin{table*}[ht]
    \centering
    \small 
    \setlength{\tabcolsep}{1.8pt} 
    \caption{Performance Comparison of Re-Ranking Methods by Category (\%)}
    \label{tab:reranking_grouped}
    \begin{tabular}{llcccccccccc c} 
        \toprule
        \multirow{2}{*}{\textbf{Category}} & \multirow{2}{*}{\textbf{Method}} & \multicolumn{3}{c}{\textbf{nDCG@$k$ (\%)}} & \multicolumn{3}{c}{\textbf{MRR@$k$ (\%)}} & \multicolumn{3}{c}{\textbf{MAP@$k$ (\%)}} & \textbf{Latency} & \textbf{Cost} \\
        \cmidrule(lr){3-5} \cmidrule(lr){6-8} \cmidrule(lr){9-11} \cmidrule(lr){12-12} \cmidrule(lr){13-13}
        & & @1 & @5 & @10 & @1 & @5 & @10 & @1 & @5 & @10 & (second/query) & (\$/1k query) \\
        \midrule
        --- & (stage 1 retrieval) & 43.2 & 45.3 & 51.1 & 43.2 & 56.0 & 57.2 & 18.2 & 36.1 & 40.1 & --- & --- \\   
        \midrule
        Late Interaction & ColBERT\cite{khattab2020colbert} & 33.6 & 31.7 & 36.7 & 33.6 & 43.3 & 45.2 & 33.6 & 41.7 & 41.0 & 7.26 & free \\
        \midrule
        \multirow{5}{*}{\shortstack[l]{Cross-\\Encoder}} 
        & BGE-reranker (base)\cite{li2023making} & 17.5 & 20.2 & 24.9 & 17.5 & 27.2 & 29.2 & 17.5 & 26.1 & 26.8 & 1.20 & free \\
        & BGE-reranker (v2-m3)\cite{chen2024bge} & 26.4 & 28.5 & 34.1 & 26.4 & 37.9 & 39.8 & 26.4 & 36.3 & 35.7 & 3.60 & free \\
        & SBERT-MiniLM (L-6-v2)\cite{reimers2019sentence} & 34.5 & 35.6 & 41.0 & 34.5 & 45.4 & 46.9 & 34.5 & 43.8 & 43.1 & 0.12 & free \\
        & Jina\_reranker (v2)\cite{jina_rerank_v2} & 30.9 & 32.1 & 37.6 & 30.9 & 39.8 & 41.1 & 30.9 & 38.2 & 36.8 & 0.59 &  0.16 \\
        & Cohere\_rerank (v3)\cite{cohere_rerank_v3} & 29.5 & 28.7 & 33.6 & 29.5 & 38.9 & 40.7 & 29.5 & 37.6 & 37.2 & 0.23 & 2.0 \\
        \midrule
        \multirow{3}{*}{\shortstack[l]{LLM as\\Re-Ranker}} 
        & LLMRerank (gpt-4o-mini)\cite{liang2022holistic} & 32.9 & 29.0 & 33.0 & 32.9 & 42.0 & 43.6 & 32.9 & 40.8 & 39.6 & 17.39 & 0.9 \\
        & LLMRerank (gpt-5.2)\cite{liang2022holistic} & 41.3 & 35.5 & 40.3 & 41.3 & 49.9 & 51.5 & \textbf{41.3} & 48.1 & 46.6 & 14.47 & 15.0 \\
        & RankGPT (gpt-4o-mini)\cite{sun2023chatgpt} & 36.3 & 42.6 & 48.0 & 36.3 & 51.1 & 52.4 & 36.3 & \textbf{49.9} & 48.8 & 23.11 & 0.4 \\
        \midrule
        \multirow{3}{*}{\shortstack[l]{Spatiotemporal-\\Aware}} 
        & GT-R\cite{tian2025advancing} & 31.1 & 33.3 & 38.9 & 31.1 & 33.3 & 38.9 & 31.1 & 33.3 & 40.4 &  1.56 & 3.3 \\
        & RRF\cite{cormack2009reciprocal} & 33.1 & 36.1 & 42.9 & 33.1 & 47.0 & 48.7 & 13.0 & 26.9 & 31.0 & 1.80 & 10.0 \\
        & \textbf{ASTRA (Ours)} & \textbf{60.7} & \textbf{57.2} & \textbf{63.8} & \textbf{60.7} & \textbf{71.1} & \textbf{72.1} & 24.7 & 46.7 & \textbf{51.8} & 1.80 & 10.0 \\
        \bottomrule
    \end{tabular}
\end{table*}

\subsubsection{Retrieval Trade-offs in Multimodal Integration}
To systematically evaluate the contribution of multimodal cues and the efficacy of our proposed architectural design, we conducted an ablation study categorized into three groups: the full CAMERA framework, text-only baselines, and visual-only baselines. The primary objective is to investigate how textual and visual modalities interact with different retrieval mechanisms (sparse vs. dense) to determine the optimal strategic alignment for high-recall event discovery.
The ablation results (Table \ref{tab:retrieval_ablation}) show that although visual cues possess independent discriminative power, achieving a 35.3\% Recall@100 in the Visual Hybrid configuration in Table \ref{tab:retrieval_ablation}(e), their effective integration requires alignment with the appropriate retrieval approach. In Group 1, the Unimodal Hybrid (b) configuration serves as a strong baseline at 93.1\%. Our proposed CAMERA (a) framework further improves this to 93.4\%. It demonstrates that CAMERA successfully extracts non-redundant contextual information from images that dense textual embeddings alone cannot capture.
To achieve this, CAMERA integrates visual cues exclusively into the dense semantic embeddings rather than the sparse index.
This design prevents the lexical noise that often occurs when descriptive visual cues are injected into sparse indexers, ensuring that the semantic depth of foundation models is maximized without compromising keyword-matching precision.
The results validate that the efficacy of GeoAI retrieval depends on coupling each modality with its most suitable retrieval engine.

\begin{table*}[ht]
\centering
\caption{Ablation Study of CAMERA Retrieval Components (Recall@100)}
\label{tab:retrieval_ablation}
\small
\begin{tabular}{l cccc cc}
\toprule
\textbf{Configuration} & \multicolumn{2}{c}{\textbf{Modality}} & \multicolumn{2}{c}{\textbf{Retrieval}} & \textbf{Recall@100} & \textbf{$\Delta$} \\
\cmidrule(lr){2-3} \cmidrule(lr){4-5}
& \textbf{Text} & \textbf{Visual} & \textbf{Sparse} & \textbf{Dense} & \textbf{(\%)} & \textbf{(\%)} \\
\midrule
\textbf{(a) CAMERA (Full)} & \checkmark & \checkmark & \checkmark & \checkmark & \textbf{93.4} & \text{--} \\
\midrule
\textit{Group 1: Text-Only} \\
(b) Hybrid        & \checkmark &            & \checkmark & \checkmark & 93.1 & $-$0.3 \\
(c) Dense         & \checkmark &            &            & \checkmark & 92.2 & $-$1.2 \\
(d) Sparse        & \checkmark &            & \checkmark &            & 86.8 & $-$6.6 \\
\midrule
\textit{Group 2: Visual-Only} \\
(e) Hybrid          &            & \checkmark & \checkmark & \checkmark & 35.3 & $-$58.1 \\
(f) Dense           &            & \checkmark &            & \checkmark & 33.2 & $-$60.2 \\
(g) Sparse          &            & \checkmark & \checkmark &            & 30.6 & $-$62.8 \\
\midrule
% \textit{Group 3: Naive Fusion} \\
% (h) Naive Multimodal Hybrid & \checkmark & \checkmark & \checkmark & \checkmark & 93.1 & $-$0.3 \\
% (i) Naive Multimodal Dense  & \checkmark & \checkmark &            & \checkmark & 93.1 & $-$0.3 \\
% (j) Naive Multimodal Sparse & \checkmark & \checkmark & \checkmark &            & 85.4 & $-$8.0 \\
% \bottomrule
\end{tabular}
\end{table*}

\subsubsection{Impact of Adaptive Features on Re-ranking}
\begin{figure}[bp]
    \centering
    \includegraphics[width=1\columnwidth]{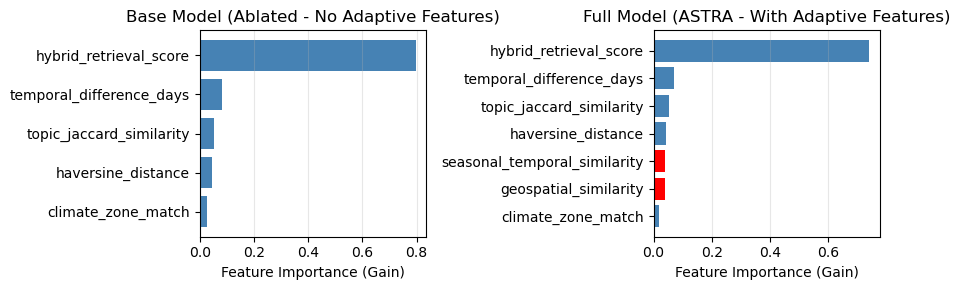} 
    \caption{Retrieval performance comparison when removing visual cues in Context-Aware Multimodal Event Retrieval Algorithm (CAMERA) for sparse, hybrid, and dense retrieval approaches.}
    \label{fig:figure-retrieval}
\end{figure}

To further investigate the contribution of the proposed ASTRA to the re-ranking process, we conducted a feature importance analysis using the \textit{Gain} metric within our LTR framework. Figure~\ref{fig:figure-retrieval}
shows a comparative view of feature contribution between the ablated base model (left) and the full ASTRA model incorporating adaptive features (right).
% \textbf{Redistribution of Feature Contribution.} 
In the base model, the ranking logic relies heavily on the hybrid retrieval score, which accounts for about 79.6\% of the importance gain. This indicates that without adaptive grounding, the re-ranker functions as a refinement of the first-stage retrieval rather than an independent evaluator of spatiotemporal context. However, adding the adaptive features (geospatial similarity and seasonal temporal similarity) improves retrieval performance, and we observe a redistribution of feature importance. The reliance on the raw retrieval score decreases.
% \textbf{Efficacy of Adaptive Spatiotemporal Grounding.} 
The emergence of seasonal temporal similarity and geospatial similarity as gain-contributing features suggests that ASTRA successfully captures multiscale relationships that traditional linear metrics (haversine distance, temporal difference days) overlook. Although the raw distance provides a static measure of proximity, our adaptive geospatial similarity allows the LTR algorithm to weight spatial relevance dynamically based on the event's specific impact scale. This shift shows that the ASTRA framework alters the ranking strategy to prioritize multiscale spatiotemporal alignment, leading to the superior performance reported in Table~\ref{tab:reranking_grouped}.

% \subsection{(Optional)Example}
% (Coming soon...)
% (adaptive scale example is added to the prompt figure instead.)

\section{DISCUSSION}

Recommending similar spatiotemporal events remains a challenging and relatively underexplored problem in GIR, particularly when event reports contain heterogeneous modalities such as text, images, and metadata. This study demonstrates that foundation models can enhance event retrieval by combining multimodal semantic understanding with geographic reasoning. The proposed framework integrates these capabilities through two complementary mechanisms, multimodal cue extraction for richer event representation and adaptive spatiotemporal reasoning for ranking. 

In particular, the CAMERA component leverages VLM to transform visual information into semantically meaningful textual cues. Beyond environmental monitoring, this cue-based multimodal representation may also be applicable to other geographic document collections where visual content provides contextual signals relevant to event interpretation.
The design of CAMERA emphasizes interpretability and geographic context extraction from visual inputs. Pretrained multimodal encoders such as CLIP provide strong image--text alignment through latent embeddings, they are optimized primarily for global semantic similarity rather than explicit reasoning about geographic attributes. By converting visual observations into structured textual cues inferred by a VLM, CAMERA allows spatial and temporal context to be explicitly incorporated into retrieval representations. This cue-based strategy inevitably trades some low-level visual detail for interpretability and contextual reasoning. 

Future work may therefore explore hybrid representations that combine cue-based semantic descriptions with latent multimodal embeddings, allowing complementary signals to jointly support event similarity estimation and balancing interpretability, robustness, and retrieval effectiveness.
Another future direction is motivated by our observation (Table 5) that environmental GIR remains heavily text-dominated. Because event reports explicitly document key entities in the text, visual cues primarily enrich or reinforce this context, yielding modest performance gains over strong text-only baselines. To unlock the full potential of multimodal GeoAI, future visual extraction should go beyond categorical cues toward capturing quantitative or implicit contexts, such as disaster severity. Furthermore, exploring agent-driven dynamic modality weighting to reduce modality redundancy is worth exploring.

Our analysis of the adaptive scale inference used in ASTRA also provides insights into the spatial reasoning capabilities of LLMs. The inferred spatial and temporal thresholds vary across events in ways that reflect contextual interpretation of geographic processes. For example, flooding events associated with regional hydrological systems may need broader spatial bandwidth, whereas localized hazards such as glacier-dammed lake outbursts produce narrower spatial bandwidth. These variations suggest that LLMs can partially reason about the geographic scale of environmental phenomena. However, the application of spatial heterogeneity is not always consistent. In anomaly scenarios such as disease outbreaks or rare biological observations, the model tends to produce uniform thresholds across locations. Similarly, for certain large-scale geological events, the reasoning process may overlook geographic variability. These observations indicate that even though LLMs demonstrate promising spatial reasoning abilities, their interpretation of geographic scale remains context-dependent and occasionally inconsistent.

Finally, several limitations and research directions are reflected on this work. Because ASTRA relies on LLM priors to infer spatial and temporal scales, it may inherit geographic biases present in the foundation models' training data \cite{manvi2024large}. Uneven geographic knowledge representation may lead to scale distortions in regions with limited data coverage. Addressing such biases may require region-aware evaluation, curated training corpora, or integration with external geographic knowledge sources. In addition, the increasing geospatial reasoning capabilities of foundation models raise important privacy considerations. Analyses of spatial reasoning processes or model outputs could potentially reveal sensitive geographic information if not carefully managed. Developing evaluation frameworks and governance mechanisms for responsible GeoAI deployment, therefore, represents an important future research direction \cite{wang2024gpt}. Although our experiments focus on environmental reports from the Local Environmental Observer Network, the proposed framework is general and can be applied to other event-centric geographic information retrieval scenarios, including disaster monitoring, scientific observations, and environmental intelligence systems.

\section{CONCLUSION}
This study introduces a novel framework for spatiotemporal event search and recommendation, integrating retrieval and re-ranking informed by the spatial and temporal reasoning capabilities of foundation models, including LLMs and VLMs. Grounded in GIScience principles of context and scale, we demonstrate that the proposed CAMERA extracts multimodal cues regarding content and context in a human-understandable way, and ASTRA generates adaptive scales in the dimensions of space and time. Our proposed framework couples them and enhances spatiotemporal event recommendation, contributing to a more comprehensive understanding of these events and their impacts. 

This research has broader implications for domains such as environmental change analysis, disaster response, and urban planning, where accurate and context-aware event retrieval is crucial for informed decision-making and effective resource allocation.
Although the application in this study focuses on the LEO Network dataset, the proposed modular framework uses on standard metadata, offering the flexibility and potential to enable intelligent event discovery systems in other critical GIR contexts. It is well-suited for disaster catalogs, situational awareness platforms, or scientific literature mining, where specific domain-based adaptation is required.

In this era of rapidly evolving large models, we hope this work sheds light on future GIR research by encouraging methodological advances that strengthen context-aware reasoning, further incorporate hybrid retrieval from multimodal information, and integrate spatial and temporal scale–dependent analysis. In this way, we can gain a better understanding of the shared and unique characteristics of geographic phenomena, helping researchers identify new patterns among similar events and connect geographically distant events through underlying relationships. Such a tool aims to support researchers in identifying and addressing new research questions, while also enhancing situational awareness of environmental change for stakeholders and the general public.

\section{ACKNOWLEDGMENT}
This research was supported in part by Google.org’s Impact Challenge for Climate Innovation Program, OpenAI’s Researcher Access Program, and the U.S. National Science Foundation (Grant Nos. 2230034 and 2120943). All event thumbnail images are credited to their original authors, as indicated on the LEO Network platform. Special thanks to Chia-Yu Hsu for technical support. The authors also acknowledge Research Computing at Arizona State University for providing the computing resources that contributed to the research results reported in this paper.

% \section{APPENDIX I}
% Prompt for Adaptive Spatiotemporal Re-ranking:
% \begin{figure}[h]
%     \centering
%     \includegraphics[width=0.8\columnwidth]{figure/appendix1.png} % Replace with your image path
%     % \caption{}
%     \label{fig:appendix1}
% \end{figure}
% \section{Appendices}
% \begin{verbatim}
%   Prompt for Adaptive Spatiotemporal Re-ranking
% \end{verbatim}

%%
%% The next two lines define the bibliography style to be used, and
%% the bibliography file.
\bibliographystyle{ACM-Reference-Format}
\bibliography{sample-base}

%%
%% If your work has an appendix, this is the place to put it.
\appendix

\end{document}